\documentclass[pre,superscriptaddress,twocolumn,nofootinbib,longbibliography]{revtex4-1}
\usepackage{amsfonts,amsmath,amssymb,graphicx,xcolor}
\usepackage[colorlinks=true,allcolors=blue]{hyperref}
\usepackage[shortlabels]{enumitem}

\begin{document}

\title{Dynamic Hidden-Variable Network Models}

\author{Harrison Hartle}
\affiliation{Network Science Institute, Northeastern University, Boston, MA, USA}

\author{Fragkiskos Papadopoulos}
\affiliation{Department of Electrical Engineering, Computer Engineering and Informatics, Cyprus University of Technology, 3036 Limassol, Cyprus}

\author{Dmitri Krioukov}
\affiliation{Network Science Institute, Northeastern University, Boston, MA, USA}
\affiliation{Northeastern University, Departments of Physics, Mathematics, and Electrical\&Computer Engineering, Boston, MA, USA}

\begin{abstract}
Models of complex networks often incorporate node-intrinsic properties abstracted as hidden variables. The probability of connections in the network is then a function of these variables. Real-world networks evolve over time, and many exhibit dynamics of node characteristics as well as of linking structure. Here we introduce and study natural temporal extensions of static hidden-variable network models with stochastic dynamics of hidden variables and links. The rates of the hidden variable dynamics and link dynamics are controlled by two parameters, and snapshots of networks in the dynamic models may or may not be equivalent to a static model, depending on the location in the parameter phase diagram. We quantify deviations from static-like behavior, and examine the level of structural persistence in the considered models. We explore temporal versions of popular static models with community structure, latent geometry, and degree-heterogeneity. We do not attempt to directly model real networks, but comment on interesting qualitative resemblances, discussing possible extensions, generalizations, and applications.
\end{abstract}

\maketitle

\section{Introduction}\label{sec:intro}
Networks are ubiquitous in nature \cite{janwa2019origin,boers2019complex,costa2011analyzing,rocha2010information,west1997general,pastor2004evolution,zheng2020geometric,cimini2019statistical,kim2019universal}, and their study relies heavily on the mathematical and computational analysis of simple models \cite{krapivsky2002statistical,barabasi1999emergence}, typically in the form of random networks built according to some stochastic rules. In many models, nodes are assigned characteristics (such as fitnesses \cite{bianconi2001competition,caldarelli2002scale} or spatial coordinates in a physical \cite{barthelemy2018morphogenesis} or latent space \cite{krioukov2010hyperbolic,serrano2012uncovering,kitsak2017latent}), which in turn affect the network's structural formation. Such models fall under the umbrella of {\it hidden-variables} models \cite{boguna2003class}, because they depend on internal node-characteristics that are only implicitly expressed by the network structure, through effects on link-formation. Usually, hidden variables (HVs) are not externally specified as parameters -- rather, their probability distribution is specified \cite{bianconi2001competition,newman2018networks}, and they are sampled during the network's formation. Two sources of randomness underly such networks: the random HVs of nodes, and the random formation of edges given those HVs. In general, hidden-variables models are defined by the following procedure:
\begin{enumerate}
\item A random hidden-variable configuration $H$ is drawn with probability density $\rho(H)$ from a set of possible hidden-variable configurations $\mathcal{H}$.
\item Graph $G$ is then drawn with conditional probability $\mathbb{P}(G|H)$ from a set of possible graphs $\mathcal{G}$.
\end{enumerate}
As a result, the overall probability of sampling any particular graph $G\in\mathcal{G}$ is equal to 
\begin{equation}\label{eq:static_graph_probability}
\mathbb{P}(G)=\int_{\mathcal{H}}\mathbb{P}(G|H)\rho(H)dH.
\end{equation}

Hidden-variables models, due to their capacity to encode nodewise heterogeneity, are in many cases capable of exhibiting more structural realism than models without hidden variables. For example, hidden variables underly network models incorporating realistic features such as community structure (stochastic block models \cite{karrer2011stochastic}), latent geometry (random geometric graphs \cite{penrose2003random}), and degree-heterogeneity (soft configuration models \cite{van2018sparse}).

However, such models do not capture the {\it dynamics} of node-characteristics, nor the impact thereof on network structure. The influence of dynamic node-states on evolving link-structure has been investigated in the context of {\it adaptive networks} \cite{risau2009contact,piankoranee2018effects,huepe2011adaptive,marceau2010adaptive,demirel2014moment,gross2009adaptive}, but in that case node-states arise due to a highly complex feedback, interacting with one another through co-evolving links. Such models are more realistic and have interesting features, but they do not directly explore the impact of dynamic node-properties on dynamic network structure. 

There is a wide abundance of real-world examples of dynamic node-properties influencing dynamics of network structure, such as:
\begin{enumerate}[a)]
\item changing habits, interests, jobs, and other attributes of people in social networks \cite{crabtree1998identifying},
\item changing geospatial coordinates of organisms during formation of social ties, group-memberships, and pathogenic contact networks \cite{chapman2011ecology,vardanis2011individuality,altizer2011animal,pfeffer2010emergence,hein2012energetic},
\item changing phenotypic traits of species as they biologically evolve in ecological networks \cite{carroll2007evolution,held2014adaptive}, 
\item changing marketing and administrative strategies of entities in economic networks \cite{volberda2003co,stoica1999understanding}, 
\item changing demographic and infrastructural characteristics of cities in evolving highway and airport networks \cite{johnson2006demographic,myers1999demographic,raimbault2018modeling},
 \item changing gene-expression levels of neurons in developing connectomes \cite{kuhar1993changing,mccormack1992changes},
 \item changing consumption-levels of residential nodes in evolving power grids \cite{dalvand2008long,de2013electricity},
 \item changing displayed content of websites on the evolving world-wide web \cite{pasichnyk2014mathematical,bean2011emerging}. 
\end{enumerate}

These examples motivate the development of a simple modeling framework describing the impact of dynamic node-characteristics on dynamic link-structure. Such a framework would provide a temporal analogue of how node-properties influence network structure in hidden-variables models. In fact, it is standard practice to derive temporal versions of static-network concepts \cite{nicosia2013graph,ortiz2017navigability,taylor2017eigenvector,kim2012temporal,pan2011path,li2017fundamental,liu2014controlling,perra2012random,paranjape2017motifs,holme2016temporal,liu2018epidemic,nadini2018epidemic,masuda2017temporal,sun2015contrasting,li2018opinion,dunlavy2011temporal,dhote2013survey,sarzynska2016null,gauvin2014detecting}, as has been done for several models of static networks with hidden variables such as stochastic block models \cite{peixoto2017modelling,xu2014dynamic,xu2015stochastic,matias2017statistical,pensky2019spectral,ghasemian2016detectability,barucca2018disentangling}.

Motivated by these considerations, here we study temporal extensions of general static hidden-variables models, obtained by introducing dynamics of hidden variables and of links. In these models each node has an evolving hidden variable, and each node-pair has a pairwise {\it affinity} (equal to the connection probability in the static hidden-variables model), which is a function of the hidden variables of both nodes. Pairwise affinities evolve over time due to their dependence on a pair of evolving hidden variables. The network itself evolves via node-pairs being selected to re-evaluate their connections, resampling them with connection probability equal to the pair's affinity at the moment of re-evaluation. These systems are governed by just two parameters beyond those of any static model: a rate of hidden-variable dynamics $\sigma$, and a rate of link-resampling $\omega$.

We find that these models have snapshots that are statistically equivalent to networks generated from the static model in the following cases: 
\begin{enumerate}[a)]
\item if there is a sufficient timescale separation (slow hidden-variable-dynamics relative to link-dynamics),
\item if connectivity is a deterministic function of hidden variables, 
\item if hidden variables are held fixed, or
\item if we add an additional dynamic mechanism whereby links actively respond to changes in hidden variables.
\end{enumerate}
We also identify the conditions under which model networks evolve {\it gradually}, {\it i.e.}, exhibit link-persistence, and evaluate qualitative resemblances of snapshots to some real networks which arise as {\it deviations} from static-model behavior. We obtain analytical and numerical results for effective connection probabilities (the probability of a node-pair being connected given their {\it current} hidden-variable values), directly quantifying deviations from static-model behavior in each case.

The family of models we introduce is demonstrated to have wide generality, as exemplified by temporal extensions of four different static models with hidden variables: stochastic block models \cite{karrer2011stochastic}, random geometric graphs \cite{penrose2003random}, soft configuration models \cite{van2018sparse}, and hyperbolic graphs \cite{krioukov2010hyperbolic}. These examples relate to, and partially encompass, several models of networks with dynamic node-properties that have been previously studied -- for instance dynamic latent space models \cite{sewell2016latent,kim2018review,sarkar2007latent,sarkar2006dynamic}, dynamic random geometric graphs \cite{peres2013mobile,clementi2015parsimonious}, and dynamic stochastic block models \cite{ghasemian2016detectability,barucca2018disentangling}. The framework we study is also widely generalizable to other contexts.

Our study takes a step towards realistic modeling of dynamic networks with dynamic node properties. It introduces a family of temporal network models that extends static hidden-variables models to the temporal setting, providing theoretical insight into the kinds of structure that can emerge as a consequence of the influence of hidden-variable dynamics on network-structure dynamics. The framework can be used for studying real-world temporal networks under the null hypothesis that physical or latent {\it dynamic} hidden variables drive the dynamics of network structure. Additionally, motivated by the phenomenology emerging in these models, we speculate that links in some real systems are {\it out of equilibrium} with respect to hidden-variables, partially explaining the presence of long-ranged links in geometrically-embedded systems and inter-group connectivity in modular systems.

In Section \ref{sec:properties}, we describe the properties that we use to characterize the models we introduce. We then introduce the static and temporal hidden-variables model families in Section \ref{sec:model}, followed by various limiting regimes in Section \ref{sec:limit_cases}. Section \ref{sec:examples} provides several examples illustrating temporal hidden-variables models. We then consider a variant of the family of models in Section \ref{sec:link_response}, incorporating an additional dynamic mechanism that enforces static-model connection probabilities. The final sections are dedicated to descriptions of related work (Section \ref{sec:related_work}) and a discussion of our results and the implications thereof (Section \ref{sec:discussion}). Appendices provide the details of several calculations and procedures left out of the main text.

\section{Desired properties of dynamic hidden-variables models}\label{sec:properties}
This section outlines the properties that we use to characterize the family of dynamic hidden-variables models that we introduce. Our goal is to construct natural temporal versions of static networks with hidden variables, and to understand the consequences of having introduced such dynamics. Our approach is via a Markov chain on graphs and hidden-variable configurations, with sources of randomness in the original static model being replaced by random {\it processes} in the temporal model. 

Specifically, given a static hidden-variables model, {\it i.e.}, a probability density on hidden-variable configurations $H\in\mathcal{H}$ and a conditional probability distribution on graphs $G\in\mathcal{G}$ given $H$, the temporal extension yields a probability distribution/density on {\it  temporal sequences} of graphs and hidden-variable configurations, denoted  $\mathbf{G}=\left\{G^{(t)}\right\}_{t=1}^T\in\mathcal{G}^T$ and  $\mathbf{H}=\left\{H^{(t)}\right\}_{t=1}^T\in\mathcal{H}^T$, respectively. We will evaluate the conditions under which models within our framework satisfy the following properties:

\begin{enumerate}[a)]\itemsep0cm
\item {\it Equilibrium Property:} The marginal probability of a graph at any timestep is identical to its probability in the static model; likewise for hidden variables.
\item {\it Persistence Property}: The level of structural persistence over time -- quantified by, e.g., any graph similarity measure between graphs at adjacent timesteps -- is high relative to the null expectation (of two i.i.d. static-model samples).
\item {\it Qualitative Realism}: The graph-structure, HV-geometry (e.g., link-lengths), and/or dynamic behaviors resemble observed characteristics of some real-world systems at a qualitative level.
\end{enumerate}

If the Equilibrium Property is satisfied, the temporal network in question is a strict extension of the static model -- individual snapshots are then indistinguishable from static-model realizations. If the Equilibrium Property is {\it not} satisfied, snapshots {\it deviate} from the static model, the resulting phenomenology of which we seek to understand. The Persistence Property holding implies a {\it gradually} evolving network, without sudden structural transitions between networks at adjacent timesteps. In most cases we have a parameter to tune the level of structural persistence, making the level of satisfaction of the Persistence Property fall along a continuum. To have Qualitative Realism simply means that the system exhibits some characteristics and behaviors that are analogous to real-world systems -- regardless of whether the detailed mechanisms are realistic or quantitatively accurate. In particular, we are interested in qualitative features relating to the dynamics of node-characteristics, and the effects of such dynamics on a network's structural evolution.

\section{Modeling framework}\label{sec:model}
This section provides an overview of our modeling approach, and then defines static and temporal hidden-variables models. We first describe our approach to constructing temporal extensions of static models, which produce length-$T$ sequences of graphs $\mathbf{G}$ with a probability conditioned on a length-$T$ sequence of hidden-variable configurations $\mathbf{H}$. The latter arises from Markovian dynamics \cite{norris1998markov,behrends2000introduction} governed by conditional probability density $\mathcal{P}_H\left(H^{(t+1)}\left\vert H^{(t)}\right)\right.$. The initial configuration $H^{(1)}$ is sampled from the static-model hidden-variable density $\rho\left(H^{(1)}\right)$. Markovian dynamics yields a temporally-joint probability density $p(\mathbf{H})$ as a product:
\begin{equation}
p(\mathbf{H})=\rho\left(H^{(1)}\right)\prod_{t=1}^{T-1}\mathcal{P}_H\left(H^{(t+1)}\left\vert H^{(t)}\right)\right..
\end{equation}
Given $\mathbf{H}$, the graph sequence $\mathbf{G}$ is produced via a Markov chain with transition probability having auxiliary $\mathbf{H}$-dependence, $\mathcal{P}_G\left(G^{(t+1)}\left\vert G^{(t)},\mathbf{H}\right)\right.$. Herein, we primarily consider graph dynamics with $\mathbf{H}$-dependence of the form $\mathcal{P}_G\left(G^{(t+1)}\left\vert G^{(t)},H^{(t+1)}\right)\right.$, but also consider dynamics of the form $\mathcal{P}_G\left(G^{(t+1)}\left\vert G^{(t)},H^{(t+1)},H^{(t)}\right)\right.$ in Section \ref{sec:link_response}. In general, we could consider any choice of $\mathbf{H}$-dependence -- as long as $G^{(t)}$ is not influenced by $H^{(t')}$ for any $t'>t$, since that would entail graph-structure at time $t$ being dependent on HVs at future-times $t'>t$. The initial graph $G^{(1)}$ is sampled from the static-model conditional probability $\mathbb{P}\left(G^{(1)}\left\vert H^{(1)}\right)\right.$. The $\mathbf{H}$-conditioned temporally-joint graph probability distribution $P(\mathbf{G}|\mathbf{H})$ is then given by:
 \begin{equation}\label{eq:P_of_g_given_h}
P(\mathbf{G}|\mathbf{H})=\mathbb{P}\left(G^{(1)}\left\vert H^{(1)}\right)\right.\prod_{t=1}^{T-1}\mathcal{P}_G\left(G^{(t+1)}\left\vert G^{(t)},\mathbf{H}\right)\right..
\end{equation}
 Altogether, the temporally-joint graph probability distribution is given by 
 \begin{equation}\label{eq:graph_sequence_probability}
P(\mathbf{G})=\int_{\mathcal{H}^{T}}P(\mathbf{G}|\mathbf{H})p(\mathbf{H})d\mathbf{H},
\end{equation}
which is the temporal extension of Equation \eqref{eq:static_graph_probability}. 

It is this strategy that underlies all temporal extensions of static models that we consider. Static graphs without hyperparameters may also be included by disregarding $\mathbf{H}$ above, leaving only Equation \eqref{eq:P_of_g_given_h}, which becomes a general Markov chain on graphs governed by $\mathcal{P}_G\left(G^{(t+1)}\left\vert G^{(t)}\right)\right.$. Note that $\mathbf{G}$ can be seen as a multiplex network \cite{bianconi2018multilayer,de2013mathematical} with layers representing timesteps.

\subsection{Static Hidden-Variables Model}\label{ssec:SHVM}
Here we describe the static hidden-variables model \cite{boguna2003class} (SHVM), which generates graphs by a two-step procedure. First, each node $j$ (out of $n$ total, labeled as $\{1,...,n\}=[n]$) is assigned a {\it hidden variable} $h_j\in\mathcal{X}$, drawn independently with probability density $\nu(h_j)$ from set $\mathcal{X}$. Thus the hidden-variable configuration is $H=\{h_j\}_{j=1}^n\in\mathcal{H}=\mathcal{X}^n$ and the joint hidden-variable density is $\rho(H)=\prod_{j=1}^n\nu(h_j)$. Second, node-pairs $ij$ $(1\le i<j\le n)$ connect with pairwise probability $f\left(h_i,h_j\right)$, independently from one another. The conditional probability $\mathbb{P}(G |H)$ of a graph $G$ is thus given by
\begin{equation}\label{eq:cond_prob}
\mathbb{P}(G|H)=\prod_{1\le i<j\le n}\left(f\left(h_{i},h_j\right)\right)^{A_{ij}}\left(1-f\left(h_i,h_j\right)\right)^{1-A_{ij}},
\end{equation}
where $\{A_{ij}\}_{1\le i<j\le n}$ are elements of the adjacency matrix of graph $G$. For a fixed $H$, this is an edge-independent random graph. But since $H$ is random, $\mathbb{P}(G)$ is a probabilistic mixture of Equation \ref{eq:cond_prob} over possible hidden-variable configurations $H\in\mathcal{X}^n$ via Equation \ref{eq:static_graph_probability}.

\subsection{Temporal Hidden-Variables Model}\label{ssec:THVM}
We now describe a temporal version of the SHVM (Section \ref{ssec:SHVM}), namely the {\it temporal hidden-variables model} (THVM). We denote by $A_{ij}^{(t)}$ the $ij$-th element of $G^{(t)}$'s adjacency matrix. The initial conditions ($G^{(1)}$, $\{h_j^{(1)}\}_{j=1}^n$) are sampled from the SHVM. For $t\in\{1,...,T-1\}$, the system updates according to:
\begin{enumerate}[a)]
 \item {\it Hidden-variable dynamics:} Each node $j$ samples $h_j^{(t+1)}$ from a conditional density $\mathcal{P}_{h}(h_j^{(t+1)}| h_j^{(t)})$, discussed below.
 \item {\it Link-resampling:} Each node-pair $ij$, with probability $\omega$, resamples $A_{ij}^{(t+1)}$ with connection probability $f(h_i^{(t+1)},h_j^{(t+1)})$. Otherwise, $A_{ij}^{(t+1)}=A_{ij}^{(t)}$.
 \end{enumerate}
Simply put, each node's hidden variable undergoes Markovian dynamics (governed by $\mathcal{P}_h$), and each node-pair $ij$ is re-evaluated for linking (with probability $\omega$ each timestep) with connection probability equal to $ij$'s current affinity-value $f(h_i^{(t+1)},h_j^{(t+1)})$. We separately consider two types of hidden-variable dynamics $\mathcal{P}_h$:
\begin{itemize}
\item[a)] {\it Jump-dynamics:} Each node $j$, with probability $\sigma\in[0,1]$, resamples its hidden-variable to obtain $h_{j}^{(t+1)}$. The conditional density for jump-dynamics is thus
\begin{equation}\label{jump_dynamics}
\mathcal{P}_{h}(h'|h)=\sigma\nu(h')+(1-\sigma)\mathbf{1}_{h}(h'),
\end{equation}
with $\mathbf{1}_{h}(h')$ being the Dirac measure.
\item[b)] {\it Walk-dynamics:} The hidden variable of every node moves to a nearby point in $\mathcal{X}$ using Brownian-like motion with the average step-length proportional to parameter $\sigma\in[0,1]$.
\end{itemize}
We implement the latter option by transforming the density $\nu(h)$ on $\mathcal{X}$ to the uniform density on $[0,1]^D$, where $D$ is the dimension of $\mathcal{X}$, using the inverse CDF transform. We then do a random walk in $[0,1]^D$, with step-size proportional to $\sigma$, preserving the uniform distribution. Transformed back to $\mathcal{X}$, the random walk increments preserve the distribution $\nu(h)$. The details are in Appendix~\ref{app:simulating_walk_dynamics}.
 
In both walk-dynamics and jump-dynamics, parameter $\sigma$ encodes the rate of change of hidden variables. Also in both cases, the transition probability density $\mathcal{P}_H$ is separable due to independence of $\{h_j^{(t)}\}_{j=1}^n$:
\begin{equation}\label{configuration_density}
\mathcal{P}_H\left.\left(H^{(t+1)}\right\vert H^{(t)}\right) = \prod_{j=1}^n\mathcal{P}_{h}\left.\left(h_j^{(t+1)}\right\vert h_j^{(t)}\right).
\end{equation}
The stationary density of the above dynamics is equal to the static-model hidden-variable density $\rho$. The density of $\mathbf{H}=\{\{h^{(t)}_j\}_{j=1}^n\}_{t=1}^T$ is also separable,
\begin{equation}
p(\mathbf{H})=\prod_{j=1}^n\left(\nu\left(h_j^{(1)}\right)\prod_{t=1}^{T-1}\mathcal{P}_{h}\left.\left(h_j^{(t+1)}\right\vert h_j^{(t)}\right)\right).
\end{equation}
The probability of a graph-sequence $\mathbf{G}$ given $\mathbf{H}$ is the temporal product \eqref{eq:P_of_g_given_h} of the following transition probabilities,
\begin{equation}
\small
\mathcal{P}_G\left(\left. G^{(t+1)}\right\vert G^{(t)},H^{(t+1)}\right) = \prod_{1\le i<j\le n}Y_{ij}^{A_{ij}^{(t+1)}}(1-Y_{ij})^{1-A_{ij}^{(t+1)}},\\
\end{equation}
with $Y_{ij}$ denoting the conditional linking probability,
\begin{equation}
\begin{aligned}
Y_{ij}&=\omega f\left(h^{(t+1)}_i,h^{(t+1)}_j\right)+(1-\omega)A_{ij}^{(t)},\\
\end{aligned}
\end{equation}
encoding the fact that link-resampling happens with probability $\omega$, and that otherwise the link (or non-link) remains the same. 

We will primarily quantify the structure of THVM snapshots via the {\it effective connection probability},
\begin{equation}\label{eq:effective_connection_probability_formula}
\bar{f}(h,h')=\lim_{t\rightarrow\infty}\mathbb{P}\left(A_{ij}^{(t)}=1\left\vert h_i^{(t)}=h,h_j^{(t)}=h'\right.\right),
\end{equation}
which, if the Equilibrium Property is satisfied, is the same as the affinity-function $f(h,h')$. If the affinity is a function of a composite variable such as the distance between or the product of the pair of hidden variables, the effective connection probability is defined analogously but for those composite quantities. We note here that the average degree (number of link-ends per node) is independent of the values of $\sigma$ and $\omega$ in THVM snapshots (see Appendix \ref{app:effective_connection_probabilities}). 

\section{Parameter space and resulting dynamics of temporal hidden variables models}\label{sec:limit_cases}
\begin{figure}
\includegraphics[scale=0.30,trim=100 0 70 0]{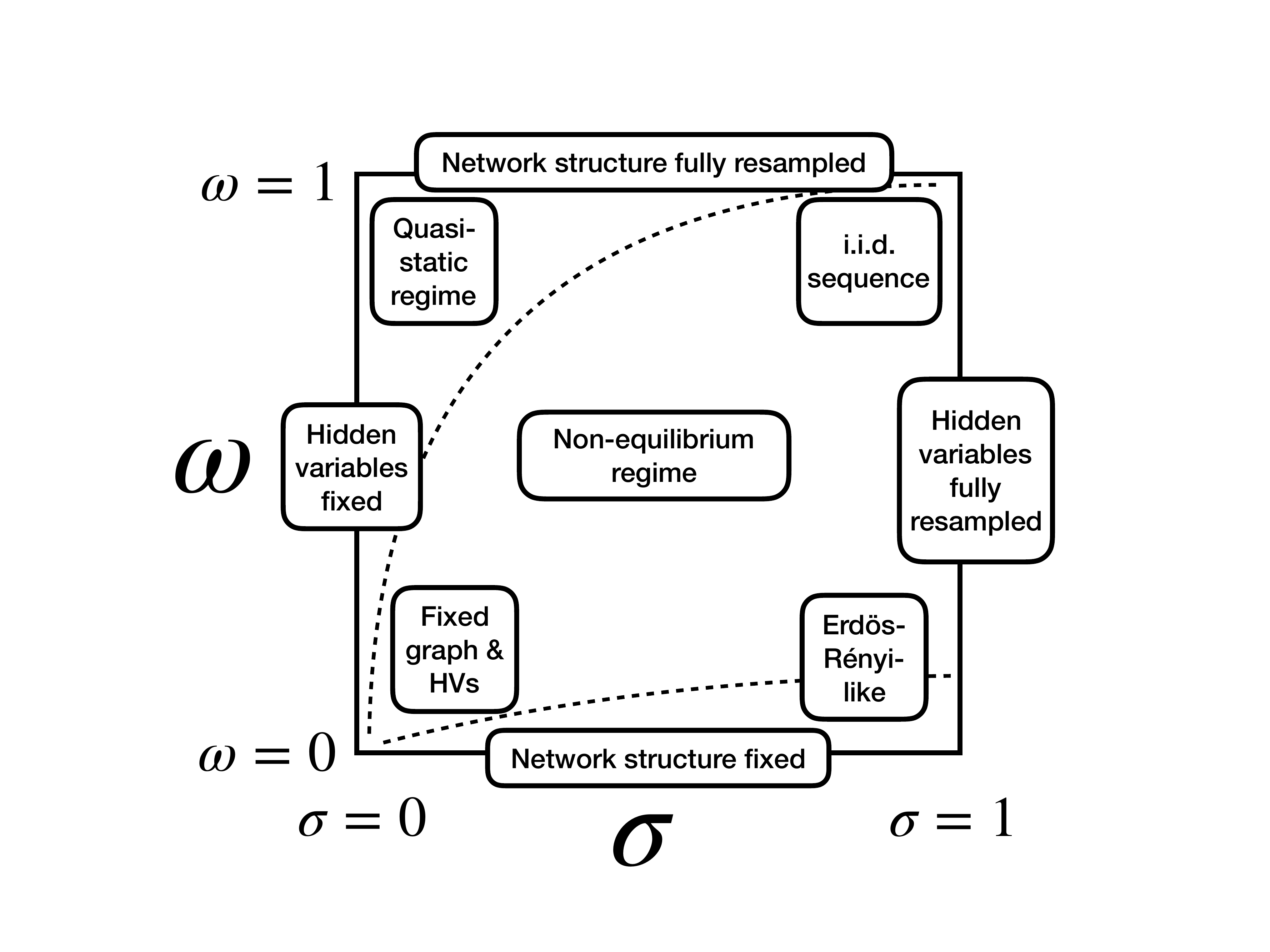}
\caption{{\bf Two-parameter space of possible dynamics.} The two parameters $(\sigma,\omega)\in[0,1]^2$ tune the rate of change of hidden variables and rate of resampling of links, respectively. In general, with dynamic hidden variables, link-structure is out-of-equilibrium relative to the configuration of hidden variables at any particular timestep, violating the Equilibrium Property. In the quasi-static regime (upper left) and along the upper and leftward boundary regions ($\omega=1$ and $\sigma=0$, respectively) the Equilibrium Property is recovered. In the lower-right regime, HVs are so randomized that network snapshots resemble Erd\H{o}s-R\'enyi graphs. At $\sigma=1$ (right-hand boundary), all hidden variables are resampled at every timestep, but only a fraction $\omega$ of links resampled. If $\omega=0$ (lower boundary), the network structure remains fixed for all time, regardless of the hidden-variable dynamics. The {\bf dashed curves} roughly designate the quasi-static regime and the Erd\H{o}s-R\'enyi-like regime.
}
\label{fig:param_space}
\end{figure}

\begin{table*}[]
\begin{tabular}{|l | l | l| l | l | }
\hline
Name & Parameter Regime & Equilibrium Property & Tunable Persistence\\
\hline
Single Static Graph & $\omega=\sigma=0$ & Yes & No\\ 
i.i.d. Graph Sequence & $\omega=\sigma=1$ & Yes & No \\
Quasi-Static & $\alpha_2(\sigma,\omega)\approx 1$ (Equation \ref{eq:alpha2}) & Yes* & Yes\\ 
Complete link-resampling & $\omega=1,\sigma\in (0,1)$ & Yes & Depends on $f$**\\
Deterministic HV-to-graph& $\omega=1, f:\mathcal{X}^2\rightarrow\{0,1\}$ & Yes & Depends on $f$**\\ 
Complete HV-resampling & $\sigma=1, \omega\in(0,1)$ & No & Yes *** \\
Fixed Hidden Variables & $\sigma=0$ & Yes & Yes \\
Erd\H{o}s-R\'enyi-like &$\sigma/\omega\gg 1$ & No & No \\
Fixed graph structure &$\omega=0$ & Yes**** & No**** \\
\hline
\end{tabular}
\caption{{\bf Table of limiting cases of dynamics-parameters $(\sigma,\omega)$ for THVMs.} The first and second columns provide a short-hand name and the associated parameter regime. The third column states whether the Equilibrium Property is satisfied, whereas the fourth column states whether the Persistence Property is satisfied (in a way that is tunable at any desired level, which for instance leaves out the case $\sigma=\omega=0$).}
\parbox{\linewidth}{\raggedright
	 \hspace{0.5cm}\scriptsize *In the quasi-static regime, $G^{(t)}$ will have arisen from an HV-configuration closely resembling $H^{(t)}$, due to a timescale-separation. This implies {\it approximate}, rather than exact, satisfaction of the Equilibrium Property.\\
	 \hspace{0.5cm}\scriptsize ** When $\omega=1$ although the persistence property is in general lost due to each possible edge being resampled at every timestep, there is still {\it some} persistence present, tuned by $\sigma$ and dependent upon the affinity function $f$. \\
	    \hspace{0.5cm}\scriptsize *** When $\sigma=1$ the persistence property is tunably satisfied at the level of graph-structure, but not at all at the level of hidden variables, which are completely resampled every timestep. \\
	     \hspace{0.5cm}\scriptsize **** In the case of $\omega=0$, the initial graph remains fixed for all time, while HVs change. Since the initial condition is sampled from the static model, this regime technically satisfies the Equilibrium Property. It does so both at the level of graphs and at the level of hidden variables, but not at all at the {\it joint} level. Persistence is not tunable at the level of graphs, but is at the level of hidden variables.
	}
\label{tab:regimes}
\end{table*}

In this section we consider several limiting cases in the space of dynamics-parameters $(\sigma,\omega)\in[0,1]^2$, and some special-case categories of affinity function $f$. The resulting regimes exhibit a variety of qualitatively distinct behaviors. If $\sigma=\omega=0$, a single graph is sampled from the static model, and all of its hidden variables and links are held fixed for all $t$. To the opposite extreme, if $\sigma=\omega=1$, at each timestep, every node's hidden variable is fully randomized, and then all possible links are re-evaluated, resulting in a sequence of independent and identically distributed (i.i.d.) instances of the static model. In either case, the Equilibrium Property is satisfied -- but the Persistence Property is not for $\sigma=\omega=1$ (there is no persistence), whereas for $\sigma=\omega=0$ there is complete persistence.

In Sections \ref{ssec:quasistatic}, \ref{ssec:complete_edge_renewal}, \ref{ssec:complete_HV_renewal}, and \ref{ssec:edge_independent}, several other parameter regimes are analyzed. We discuss the behavior of temporal networks in each case, how well they qualify in terms of the Equilibrium and Persistence Properties, and their relations to preexisting commonly studied static network ensembles. Table \ref{tab:regimes} shows the different special cases, while a schematic picture of the space of dynamics-parameters is shown in Figure \ref{fig:param_space}.

\subsection{Quasi-Static Regime ($\alpha_2(\sigma,\omega)\approx 1$)}\label{ssec:quasistatic}
Here we consider the parameter regime quantified by the condition $\alpha_2(\sigma,\omega)\approx 1$ (upper-left region of Figure \ref{fig:param_space}), where
\begin{equation}\label{eq:alpha2}
\alpha_2(\sigma,\omega)=\frac{\omega}{1-(1-\omega)(1-\sigma)^2}\in[0,1],
\end{equation}
in which networks have both random link-structure and random hidden variables, and exhibit both the Persistence Property and the Equilibrium Property. The quantity $\alpha_2(\sigma,\omega)$ is a naturally-arising function characterizing how effective connection probabilities differ from their static-model counterparts (see Appendix \ref{app:effective_connection_probabilities}). The Equilibrium Property is satisfied due to sufficient timescale separation: link-resampling happens quickly enough relative to hidden-variable motion for $G^{(t)}$ to remain caught up with $H^{(t)}$. The dynamics can thus be considered quasi-static, in the sense of quasi-static transformations in classical equilibrium thermodynamics \cite{landsberg1956foundations}. Over time, the HV-configuration and link-structure both fully explore their respective spaces, functioning as a temporal network whose stationary distribution is the static hidden-variables model defined in Section \ref{ssec:SHVM}. Note that the Equilibrium Property is only {\it approximately} satisfied if $\alpha_2(\sigma,\omega)<1$, that approximation becoming exact only in limit of extreme timescale-separation or $\alpha_2(\sigma,\omega)=1$. Two regimes at the boundary of the quasi-static regime have exact satisfaction of the Equilibrium Property: $\omega=1$ (Section \ref{ssec:complete_edge_renewal}) and $\sigma=0$ (Section \ref{ssec:edge_independent}). Adding a third mechanism of dynamics allows for exact satisfaction of the Equilibrium Property at all $(\sigma,\omega)\in[0,1]^2$ (see Section \ref{sec:link_response}).

\subsection{Complete link-resampling ($\omega=1$)}\label{ssec:complete_edge_renewal}
Here we consider the case $\omega=1$ (top region of Figure \ref{fig:param_space}). This case resembles that of the quasi-static regime, but all links form based on {\it current} hidden-variable configurations, so there is no graph-encoded memory: $\mathcal{P}_G\left(G^{(t+1)}\left\vert G^{(t)},\mathbf{H}\right)\right.=\mathbb{P}\left(G^{(t+1)}\left\vert H^{(t+1)}\right)\right.$. The resulting Markov chain on $\mathcal{H}\times\mathcal{G}$ thus satisfies the Equilibrium Property {\it exactly}, as opposed to approximately in the quasi-static regime (Subsection \ref{ssec:quasistatic}). Link-structure when $\omega=1$ is more correlated over time than two i.i.d. samples from the SHVM (due to persistence in HV-configurations), but the specific level of persistence depends on the form of the affinity function $f(h,h')$ and on $\sigma$. A variety of temporal network models have fully-resampled edges at each timestep \cite{perra2012activity,papadopoulos2019latent,ghasemian2016detectability}.

As subset of the $\omega=1$ regime, consider THVMs with binary affinity function $f:\mathcal{X}^2\rightarrow\{0,1\}$. In this case {\it all randomness comes from hidden variables}, because $f$ deterministically maps HV-configurations to graphs. The static model's conditional probability distribution in such cases is given by a product of indicator functions:
\begin{equation}
\mathbb{P}(G|H)=\prod_{1\le i<j\le n}\mathbf{1}\left\{A_{ij}=f\left(h_i,h_j\right)\right\},
\end{equation}
equal to $1$ if and only if $f(h_i,h_j)=A_{ij}$ for all $ij$, and equal to zero otherwise. Since the HV-dynamics $\mathcal{P}_H$ conserves $\rho$, and since $\omega=1$ ensures that all node-pairs have up-to-date links with respect to hidden variables, this model satisfies the Equilibrium Property {\it exactly}. The rate of HV-dynamics, and thus of link-dynamics, is controlled by $\sigma$ (but also influenced by the form of $f$). This regime encompasses sharp random geometric graphs (RGGs) of any kind \cite{penrose2003random}; see Section \ref{ssec:temporal_random_geometric} for temporal RGGs with $\omega\in[0,1]$.

\subsection{Fixed Hidden Variables ($\sigma=0$)}\label{ssec:edge_independent}
Here we consider $\sigma=0$ (left region of Figure \ref{fig:param_space}), in which case all HVs are frozen in place, ensuring satisfaction of the Equilibrium Property. The initial HV-configuration $H^{(1)}$ has the SHVM density $\rho$, but conditioning on some particular initial configuration $H^{(1)}$ yields fixed pairwise connection probabilities $p_{ij}=f\left(h_i,h_j\right)$, resulting in temporal versions of edge-independent static networks \cite{bollobas2007phase,oliveira2009concentration,lu2012spectra}. Analytical expressions for link-dynamics can be written straightforwardly in terms of the set of values $\{p_{ij}\}_{1\le i<j\le n}$ and the parameter $\omega$. The transition probability $\mathcal{P}_{G}\left(G^{(t+1)}\left\vert G^{(t)}\right)\right.$ is
\begin{equation}
\mathcal{P}_G(G^{(t+1)}|G^{(t)})=\prod_{1\le i<j\le n}p^{A_{ij}^{(t)}\rightarrow A_{ij}^{(t+1)}}_{ij},
\end{equation}
where $p_{ij}^{0\rightarrow 0}$, $p_{ij}^{0\rightarrow 1}$, $p_{ij}^{1\rightarrow 0}$, and $p_{ij}^{1\rightarrow 1}$ are respectively the non-link persistence, link-formation, link-removal, and link-persistence probabilities for node-pair $ij$. That is, $p^{\alpha\rightarrow\beta}_{ij}=\mathbb{P}(A_{ij}^{(t+1)}=\beta|A_{ij}^{(t)}=\alpha)$, given by: 
\begin{equation}\begin{aligned}\label{eq:link_transition_probability}
p_{ij}^{\alpha\rightarrow \beta}=&(1-\omega p_{ij})^{(1-\alpha)(1-\beta)}(\omega p_{ij})^{(1-\alpha)\beta}\\
&\times(\omega(1-p_{ij}))^{\alpha(1-\beta)}(1-\omega(1-p_{ij}))^{\alpha\beta}.
\end{aligned}\end{equation}

Many static network models have independent edges with pre-defined connection probabilities, and thus can be made temporal as THVMs with $\sigma=0$. Examples include the Erd\H{o}s-R\'enyi (ER) model \cite{erdos1959random} the (soft) stochastic block model (SBM) \cite{anderson1992building}, and inhomogeneous random graphs \cite{bollobas2007phase} with fixed coordinates.

The Persistence Property can be quantified by any of the numerous measures of graph dissimilarity \cite{hartle2020network}, by application to graph-pairs at neighboring timesteps. A simple example in the $\sigma=0$ setting is the expected Hamming dissimilarity \cite{hamming1950error},
\begin{equation}\begin{aligned}
\sum_{1\le i<j\le n}&\mathbb{P}\left(A_{ij}^{(t)}\ne A_{ij}^{(t+1)}\right)\\
&=\sum_{1\le i<j\le n}\left(p_{ij}^{1\rightarrow 0}p_{ij}+p_{ij}^{0\rightarrow 1}(1-p_{ij})\right)\\
&=2\omega \sum_{1\le i<j\le n}p_{ij}(1-p_{ij}),
\end{aligned}\end{equation}
which simplifies substantially in some cases, for instance the ER model ($p_{ij}=p$ for all $ij$), leaving $2\omega p(1-p)\binom{n}{2}$.

Edge-resampling dynamics with fixed $p_{ij}$-values closely resembles {\it dynamic percolation} \cite{steif2009survey}, which has been investigated in lattices \cite{peres1998number}, trees \cite{khoshnevisan2008dynamical} and ER graphs \cite{roberts2018exceptional,rossignol2020scaling}, and also relates to {\it edge-Markovian networks} \cite{clementi2010flooding,whitbeck2011performance,de2013relevance}.

\subsection{Complete resampling of hidden variables ($\sigma=1$)}\label{ssec:complete_HV_renewal}
Here we consider the case for which all hidden variables are resampled at every timestep ($\sigma=1$), so that no HV-driven structural persistence exists (right region of Figure \ref{fig:param_space}). Note that walk-dynamics is parameterized by $\sigma$ so that $\sigma=1$ implies complete HV-randomization. If $\sigma=1$, correlations among links (and non-links) do still exist due to simultaneous resampling; the set of node-pairs selected for link-resampling at timestep $t$ form links based upon the {\it same} underlying hidden-variable configuration $H^{(t)}$. In this setting, $\omega$  quantifies the level of agreement among node-pairs as to what the HV-configuration is. For instance in spatial network models, if $\sigma=1$, then $\omega$ directly controls the level of geometry-induced correlations.

Given the HV-configuration at time $t$ and averaging over all past timesteps, node-pair $ij$ is connected with probability
\begin{equation}\label{eq:complete_HV_renewal_effective}
\mathbb{P}\left(A_{ij}^{(t)}=1\left\vert h_i^{(t)},h_j^{(t)}\right.\right)=\omega f\left(h_i^{(t)},h_j^{(t)}\right)+(1-\omega)\langle f\rangle,
\end{equation}
where $\langle f\rangle$ is the expected affinity of a pair of nodes with randomized HVs, 
\begin{equation}
\langle f\rangle=\int_{\mathcal{X}^2}\nu(h)\nu(h')f(h,h')dhdh'.
\end{equation}
The expression \ref{eq:complete_HV_renewal_effective} is an example of an effective connection probability which deviates from the static-model affinity function. A more general formula for the effective connection probability in the case of jump-dynamics and arbitrary $f(h,h')$, $\sigma$, and $\omega$ is derived in Appendix \ref{app:effective_connection_probabilities}, and some special cases are described in the examples in Sections \ref{ssec:temporal_stochastic_block},\ref{ssec:temporal_random_geometric},\ref{ssec:temporal_hypersoft_configuration},\ref{ssec:temporal_hyperbolic_graphs}. As $\omega\rightarrow 0$ with $\sigma=1$ (and in general for $\sigma/\omega\gg 1$), the model approaches a temporal version of the ER model, since each node-pair at the time of link-resampling will have completely randomized hidden variables; each edge will then independently exist with probability $\langle f\rangle$ if $0<\omega\ll 1$. If $\omega=0$ we have fixed graph structure, {\it i.e.} a network that simply remains as whatever the initially sampled graph was, but with dynamic hidden variables (for any $\sigma>0$).

\section{Temporal extensions of popular static network models}\label{sec:examples}
This section contains several examples of THVMs. In each subsection, we describe a static hidden-variables model, its temporal extension according to the modeling framework of Section \ref{ssec:THVM}, the effective connection probability that arises due to the dynamics, and offer some additional discussion. We specifically consider temporal extensions of the following static network models: stochastic block models \cite{karrer2011stochastic}, random geometric graphs \cite{penrose2003random}, hypersoft configuration models \cite{van2018sparse}, and hyperbolic graphs \cite{krioukov2010hyperbolic}.

\subsection{Temporal Stochastic Block Models}
\label{ssec:temporal_stochastic_block}
This subsection considers temporal extensions of stochastic block models (SBMs), which are used to model community structure in networks \cite{anderson1992building,karrer2011stochastic,peixoto2012entropy,peixoto2017nonparametric}. 

\begin{figure*}
\includegraphics[trim=0 50 10 0, scale=0.5]{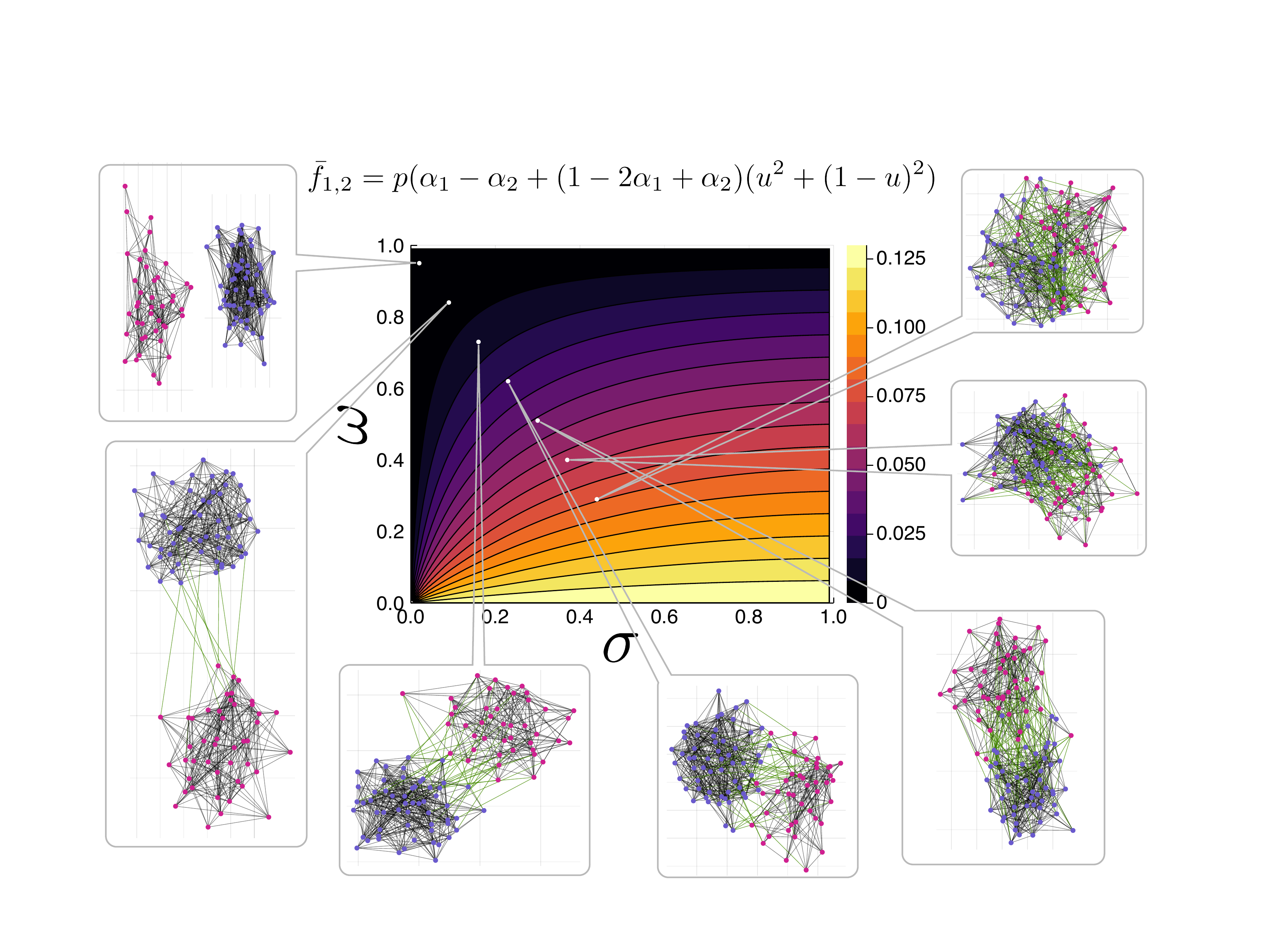}
\caption{{\bf Snapshots of a temporal stochastic block model: a modular network with dynamic group-assignments and link-resampling.} The $n=100$ nodes are partitioned into two groups with group-membership probabilities $\varrho_1=0.4=1-\varrho_2$, and group-memberships change in time by group-resampling with probability $\sigma$. The affinity function is $f_{q,q'}=p\mathbf{1}\{q=q'\}$ with $p=0.25$, disallowing inter-group connections in the {\it static} model. Network snapshots are displayed via a spring-force layout algorithm \cite{kobourov2012spring}, for various parameters $(\sigma,\omega)$ such that networks span a variety of structural outcomes. {\bf Node-coloration} is by group-membership, and {\bf link-coloration} is black for within-group links and green for between-group links. In the {\bf central panel}, the effective connection probability $\bar{f}_{1,2}$ between communities is plotted. Outside of the quasi-static regime, group-membership dynamics is fast enough for a substantial number of inter-group links to exist ($\bar{f}_{1,2}>0$), despite the inter-group connection {\it formation} probability being $f_{1,2}=0$.}
\label{fig:TSBM_phase}
\end{figure*}

\subsubsection{Static Hyperparametric SBMs}
We consider a static network with conditionally Bernoulli-distributed edges amongst $n$ nodes $j\in[n]$, each node having been randomly assigned to one of $m$ groups (a.k.a. communities, blocks, colors). Each node $j$ independently draws a group-index $q_j\in[m]=\{1,...,m\}$ from probability distribution $\varrho=\{\varrho_q\}_{q\in[m]}$. Each node-pair then connects with probability $f_{q_i,q_j}$. In this definition, the group-memberships $\{q_j\}_{j\in[n]}$ are not externally specified as model parameters -- rather, their distribution $\varrho$ is specified. Thus, the group-memberships are hyperparameters, and we refer to these static networks as hyperparameteric SBMs or hyper-SBMs (equivalent to inhomogeneous random graphs with hidden color \cite{soderberg2003random,soderberg2002general}). The expected number of nodes $n_q$ in a given block $q$ is $\langle n_q\rangle=n\varrho_q$, and the joint distribution of $\{n_q\}_{q\in[m]}$ is multinomial. Note that this model could be formulated with continuous HVs as per Section \ref{ssec:SHVM}, but we instead use discrete HVs for simplicity (see Appendix \ref{app:discrete_hidden_variables} for the continuous-to-discrete mapping). As an illustrative example to be used throughout this section, we consider the case of $m=2$ groups, with $\varrho_1=1-\varrho_2=u$. The within-group affinity is $p=f_{1,1}=f_{2,2}$, and the between-group affinity is zero ($f_{1,2}=0$).

\subsubsection{Temporal hyper-SBMs}
To make the hyper-SBM dynamic, at each timestep $t\in\{2,...,T\}$ each node $i$ with probability $\sigma$ resamples its group-index $q_i^{(t)}$ from distribution $\varrho$, and then each node-pair $ij$ with probability $\omega$ resamples $A_{ij}^{(t)}$ with connection probability $f_{q_i^{(t)},q_j^{(t)}}$. Thus,
\begin{equation}
    \mathbb{P}\left(q_i^{(t)}=q'\left\vert q_{i}^{(t-1)}=q\right.\right)=(1-\sigma)\mathbf{1}\{q=q'\}+\sigma\varrho_{q'},
\end{equation}
and
\begin{equation}
    \mathbb{P}\left(A_{ij}^{(t)}=1\left\vert A_{ij}^{(t-1)},q_i^{(t)},q_j^{(t)}\right.\right)=(1-\omega)A_{ij}^{(t-1)}+\omega f_{q_i^{(t)},q_j^{(t)}}.
\end{equation}
See Figure \ref{fig:TSBM_phase} for visualized embeddings of network snapshots from the stationary distribution of the example $m=2$, $\varrho_1=u=1-\varrho_2$, $f_{q,q'}=p\mathbf{1}\{q=q'\}$.

\subsubsection{Effective connection probabilities in hyper-SBMs}
\label{sssec:tsbm_effective}
The block-dynamics of nodes in temporal hyper-SBMs introduces several novel features to the system. First, pairwise affinities change over time. Second, the set of all existing links at time $t$ need not have arisen from the group-assignments of time $t$. Temporal snapshots in general thus deviate from the static model -- the Equilibrium Property does not necessarily hold. However, even if snapshots do not resemble {\it the} static model, they do resemble {\it a} static model -- an {\it effective} SBM. Consider two nodes, with current group-indices $q,q'$. Averaging over all past values of hidden variables, we obtain the effective connection probability $\bar{f}_{q,q'}$ for dynamic hyper-SBMs. Since the SBM case is directly obtainable from discretization of the continuous model (see Appendix \ref{app:discrete_hidden_variables}) we can use a discrete version of the general formula derived in Appendix \ref{app:effective_connection_probabilities}, namely:
\begin{equation}\begin{aligned}\label{eq:TSBM_effective_connection}
\bar{f}_{q,q'} &=  \alpha_2 f_{q,q'}\\
& \ \ +(\alpha_1-\alpha_2)\left(\langle f_{q,\cdot}\rangle+\langle f_{q',\cdot}\rangle\right)\\
&\ \ +(1-2\alpha_1+\alpha_2)\langle f\rangle,
\end{aligned}
\end{equation}
where coefficients $\alpha_b(\sigma,\omega)$ for $b\in\{1,2\}$ are given by
\begin{equation}\label{eq:alpha_param}
\alpha_b=\frac{\omega}{1-(1-\omega)(1-\sigma)^b},
\end{equation}
and marginally-averaged affinities are
\begin{equation}\begin{aligned}
\langle f_{q,\cdot}\rangle &= \sum_{q'}\varrho_{q'}f_{q,q'},\\
\langle f\rangle &= \sum_{q}\varrho_q\langle f_{q,\cdot}\rangle=\sum_{q,q'}\varrho_{q}\varrho_{q'}f_{q,q'}.
\end{aligned}\end{equation}

Note that when $\sigma=1$ we have $\alpha_1(1,\omega)=\alpha_2(1,\omega)=\omega$ and Equation \ref{eq:TSBM_effective_connection} reduces to the form of Equation \ref{eq:complete_HV_renewal_effective}. In the simple example case ($m=2,\varrho_1=u,f_{q,q'}=p\mathbf{1}\{q=q'\}$), terms in $\bar{f}_{q,q'}$ are evaluated as:
\begin{equation}
\begin{aligned}
\langle f_{1,\cdot}\rangle &= up,\\
\langle f_{2,\cdot}\rangle &= (1-u)p,\\
\langle f\rangle &=p(u^2+(1-u)^2),\\
\end{aligned}
\end{equation}
from which the formula for $\bar{f}_{q,q'}$ becomes
\begin{equation}\begin{aligned}
\bar{f}_{q,q'}& = \alpha_2 p\mathbf{1}\{q=q'\}\\
& \ \ +(\alpha_1-\alpha_2)\left\{\begin{array}{cc} 2up,& q=q'=1\\2(1-u)p, & q=q'=2\\ p & q\ne q' \end{array}\right.\\
& \ \ +(1-2\alpha_1+\alpha_2)p(u^2+(1-u)^2).
\end{aligned}
\end{equation}
In particular, the between-group effective connection probability becomes
\begin{equation}
\bar{f}_{1,2}=p\left(\alpha_1-\alpha_2+(1-2\alpha_1+\alpha_2)(u^2+(1-u)^2)\right),
\end{equation}
which is visualized in Figure \ref{fig:TSBM_phase}. In the extreme case of $\sigma/\omega\gg 1$ all links form between nodes with effectively random group-assignments, making all pairs equally likely to connect, and reducing the system to a temporal Erd\H{o}s-R\'enyi network of connection probability $p(u^2+(1-u)^2)$. 

\subsubsection{Temporal hyper-SBMs discussion}
Interesting examples of Qualitative Realism arise in temporal hyper-SBMs. For instance, group-dynamics of nodes yields inter-group connectivity, as is observed in real systems. If someone joins a different club, switches political party, or emigrates to a new country, they at first primarily carry ties to their original group -- and thus upon changing group-membership, they suddenly have many inter-group links -- {\it not} because of inter-group link-formation, but because of dynamic group-membership. Likewise, within-group connectivity can be {\it lower} than in the static model, as is the case in real systems due to nodes having recently arrived from another group, or from neighbor-nodes having recently departed. These effects arise {\it outside} the quasi-static regime, so we speculate that in some cases the {\it non-equilibrium} regime can better emulate real-world systems. We also note that we here considered group-resampling HV-dynamics (a discrete version of jump-dynamics), but we could also consider a general Markov chain on group-assignments with stationary distribution $\varrho$.

\subsection{Temporal Random Geometric Graphs}
\label{ssec:temporal_random_geometric}
In this section we describe THVMs arising from static random geometric graphs (RGGs), which model the influence of an underlying geometry on graph-structure \cite{penrose2003random}.

\begin{figure*}
\includegraphics[scale=0.6,trim=100 100 20 0]{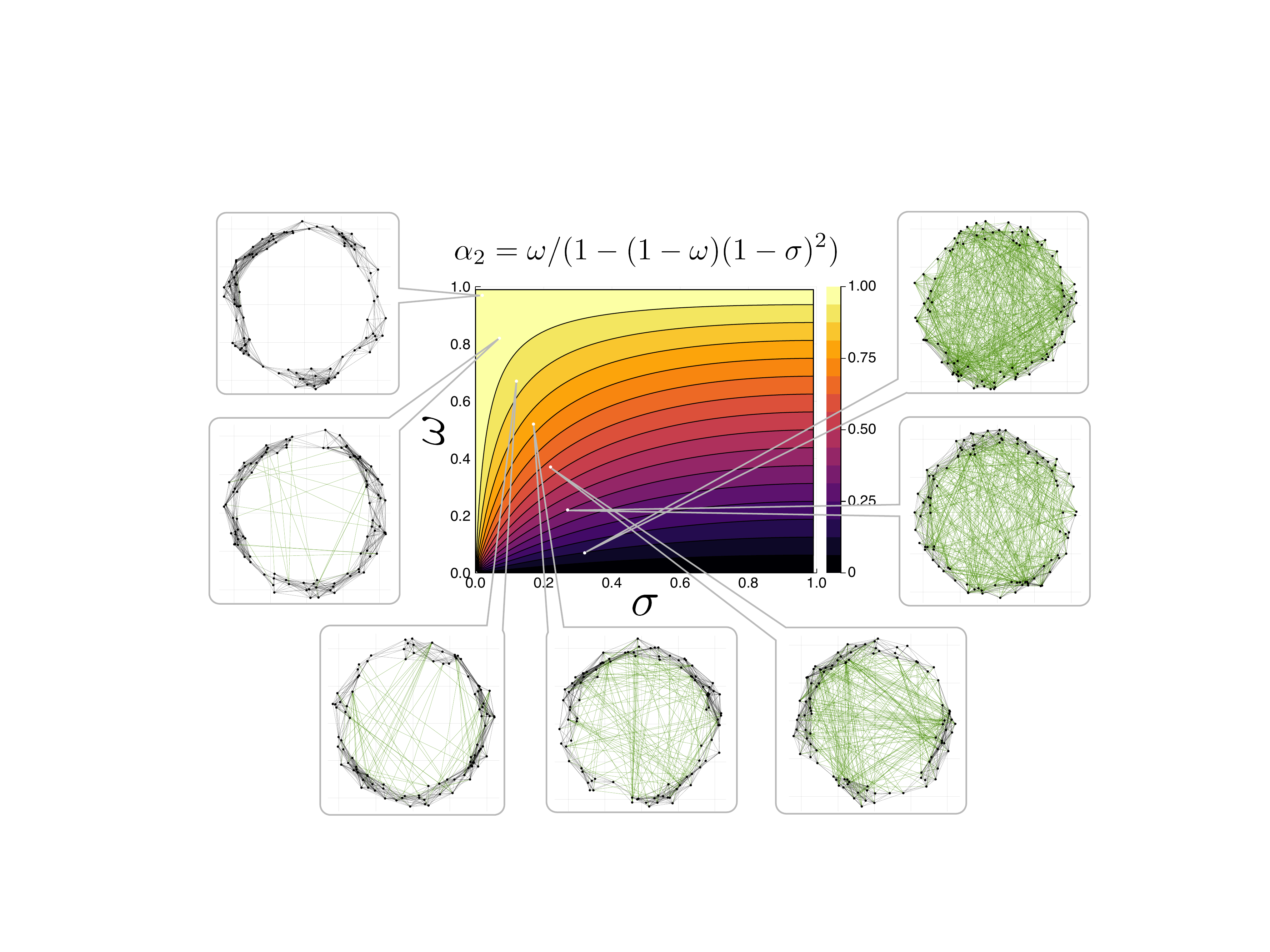}
\caption{{\bf Snapshots of a temporal random geometric graphs: a geometrically-embedded network with dynamic node-coordinates and link-resampling.} Coordinates of $n=100$ nodes are sprinkled uniformly into a 1D ring of unit circumference, and change in time via jump-dynamics (coordinate-resampling with probability $\sigma$). The affinity as a function of distance is $f(x)=\mathbf{1}\{x\le r\}$, where $r=0.1$ is the connection radius, disallowing long-ranged links in the {\it static} model. Snapshots are shown at various values of $(\sigma,\omega)$, with angular positions equal to $2\pi$ times spatial coordinates, and radial positions set to $1$ with some added random noise. {\bf Link coloration} is according to length:  black links are of distances $x\le r$ whereas green links are of distances $x>r$. In the {\bf central panel}, the function $\alpha_2(\sigma,\omega)\in[0,1]$ is visualized, which encodes the level of locality in temporal RGGs (see Equation \ref{eq:rgg_effective_connection_probability}). }
\label{fig:RGG_composite}
\end{figure*}

\begin{figure} 
\includegraphics[scale=0.25,trim=20 0 0 0]{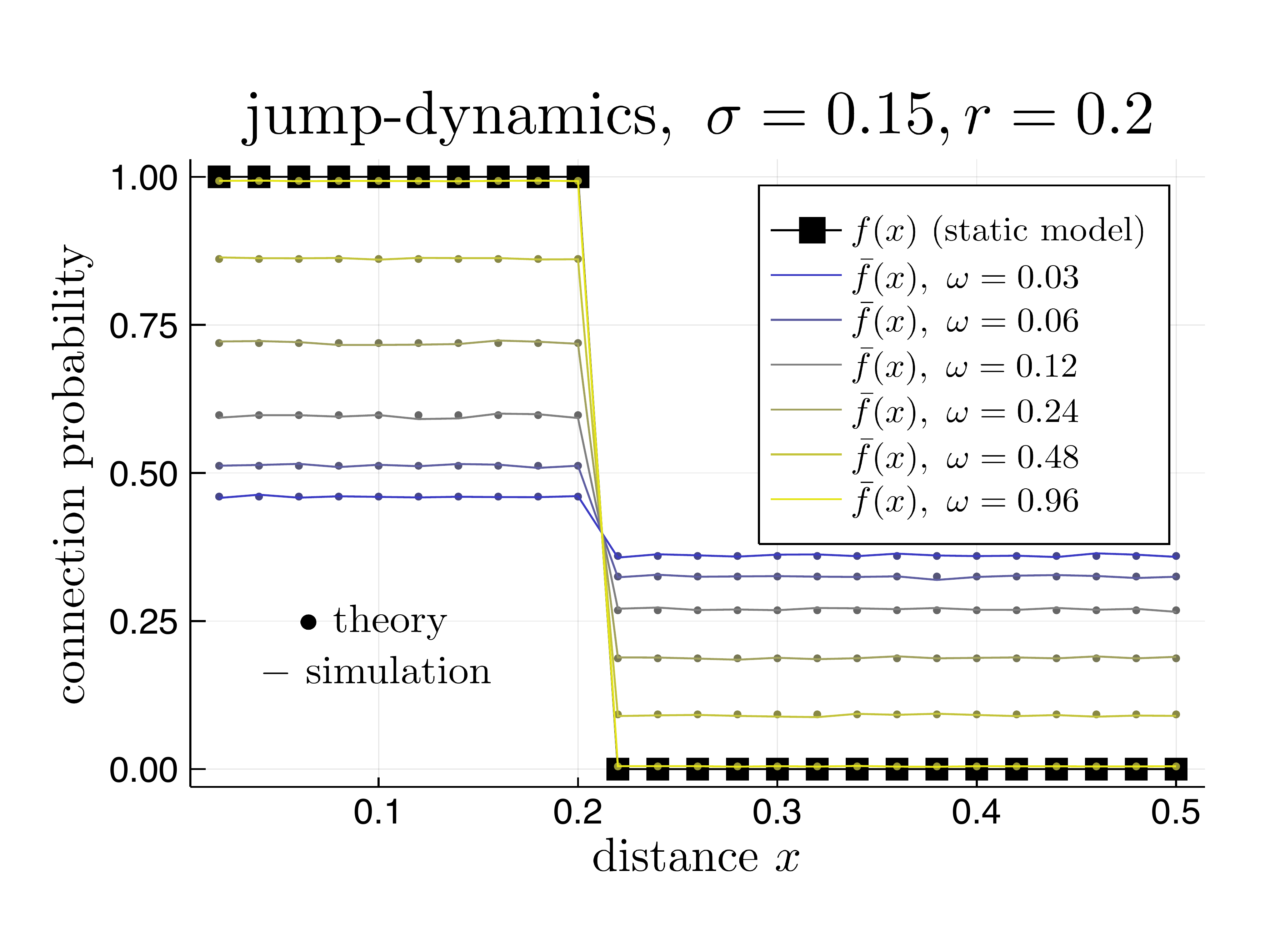}
\caption{{\bf The effective connection probability, in theory and simulation, for 1D RGGs at various values of the dynamics-parameters $(\sigma,\omega)$.} The static model affinity-function $f(x)$ is plotted with {\bf square markers}. The {\bf solid lines} are numerical estimates of the effective connection probability $\bar{f}(x)$ (with $\omega$ {\bf increasing as colors change from blue to yellow}), whereas the {\bf dotted lines} are the theoretical effective connection probability (Equation \ref{eq:rgg_effective_connection_probability}). }
\label{fig:RGG_effective}
\end{figure}

\subsubsection{Static Random Geometric Graphs}
\label{sssec:rgg}
In random geometric graphs (RGGs), nodes are assigned spatial coordinates as hidden variables, and node-pairs are linked if their coordinates are closer than some threshold distance $r$. Hence the affinity is binary, $f(h_i,h_j)=\mathbf{1}\{\mathrm{d}_{\mathcal{X}}(h_i,h_j)\le r\}$, with $\mathrm{d}_{\mathcal{X}}:\mathcal{X}^2\rightarrow[0,\infty)$ denoting the geodesic distance in latent space $\mathcal{X}$. Examples of well-studied RGGs include Euclidean RGGs with periodic or nonperiodic boundary conditions \cite{penrose2003random}, spherical RGGs \cite{allen2018random}, and hyperbolic RGGs (the hyperbolic model with inverse-temperature parameter $\beta=\infty$ \cite{krioukov2010hyperbolic}).  As a primary example we consider a simple one-dimensional RGG with periodic boundary conditions: $\mathcal{X}=[0,1)$ and $\mathrm{d}_{\mathcal{X}}(h_i,h_j)=1/2-|1/2-|h_i-h_j||$.

\subsubsection{Temporal RGGs}
To go from static RGGs to temporal RGGs, we incorporate coordinate-dynamics and link-resampling dynamics. We consider here jump-dynamics, each node resampling its coordinate according to the static-model density $\nu$, with probability $\sigma$, each timestep $t\in\{2,...,T\}$ (the coordinate density follows Equation \ref{jump_dynamics}, with $\nu(h)=1$ for the uniform density on the unit interval). Link-resampling happens independently for each node-pair with probability $\omega$ each timestep. Since RGGs have deterministic connectivity, link-resampling of $ij$ at time $t$ guarantees that $A_{ij}^{(t)}=1$ if $\mathrm{d}_{\mathcal{X}}(h_i^{(t)},h_j^{(t)})\le r$ and $A_{ij}^{(t)}=0$ otherwise. But if $ij$'s connectivity is not resampled at time $t$, links may fall out-of-equilibrium with respect to coordinates. Note that we could also study temporal RGGs with walk-dynamics, with either periodic or reflecting boundary conditions; for simplicity, we study jump-dynamics here, leaving temporal RGGs with walk-dynamics for a future study.

\subsubsection{Effective connection probabilities in temporal RGGs}
\label{sssec:trgg_effective}
We now describe the effective connection probability $\bar{f}(x)$ for RGGs between pairs of nodes for arbitrary $(\sigma,\omega)$. The expression for $\bar{f}(x)$ in temporal RGGs is derived in Appendix \ref{app:temporal_rgg_effective}, and the result is provided here:
\begin{equation}\label{eq:rgg_effective_connection_probability}
\bar{f}(x)=\alpha_2\mathbf{1}\{x\le r\}+2r(1-\alpha_2).
\end{equation}
The quantity $\alpha_2=\alpha_2(\sigma,\omega)$, defined in Equation \ref{eq:alpha_param}, directly governs the level of locality in temporal RGGs. See Figure \ref{fig:RGG_composite} for a visualization of the function $\alpha_2(\sigma,\omega)$ and of network snapshots across a range of $(\sigma,\omega)$-values. The effective connection probability $\bar{f}(x)$ has a step-like form, with connection probability $\alpha_2+2r(1-\alpha_2)$ for all $x\le r$ and $2r(1-\alpha_2)$ for all $x>r$. The above effective connection probability agrees perfectly with the results of numerical simulations, see Figure \ref{fig:RGG_effective}. 

\subsubsection{Temporal RGGs discussion}
The naturally arising function $\alpha_2(\sigma,\omega)\in[0,1]$ describes the level of locality in network snapshots (see Figure \ref{fig:RGG_composite}), and quantifies the Equilibrium Property. It interpolates between the case of RGGs ($\alpha_2(\sigma,\omega)=1$) and ER graphs ($\alpha_2(\sigma,\omega)=0$), resembling the structural transition of the Watts-Strogatz model \cite{watts1998collective}. In this case, all links {\it form} locally, and it is dynamics of {\it node positions} that induces the transition (alongside formation of local links at nodes' new locations); a similar phenomenon has been observed in contagion-dynamics among mobile agents \cite{buscarino2008disease}. Also note, in dynamic RGGs, links can exist that were not {\it possible} in the static model model: links of length greater than $r$, since the effective connection probability $\bar{f}(x)$ no longer goes completely to zero for $x>r$ (see Equation \ref{eq:rgg_effective_connection_probability}). This is related to phenomena observed in real-world networks: pairs of people may form friendships locally, but maintain those friendships after becoming geographically separated, resulting in the existence of long-ranged social ties that would not likely have {\it formed} at that distance. Likewise, the function $\bar{f}(x)$ is also {\it less than one} for distances $x\le r$, allowing for {\it non-links} that would be impossible in the static model. That phenomenon also appears in real-world systems: instead of individuals knowing everyone in their local vicinity, non-links between closeby pairs may exist, due to them having only recently become proximate. As with the case of temporal hyper-SBMs, these examples of Qualitative Realism are {\it in conflict} with the Equilibrium Property. Note also that similar deviations of $\bar{f}(x)$ relative to $f(x)$ occur in THVMs arising from {\it soft} random geometric graphs \cite{penrose2016connectivity,wilsher2020connectivity,kaiser2004spatial,dettmann2016random}, for example the $\mathbb{H}^2$ model (see Section \ref{ssec:temporal_hyperbolic_graphs}).

\subsection{Temporal Hypersoft Configuration Model}
\label{ssec:temporal_hypersoft_configuration}
In this section we consider a dynamic version of hypersoft configuration models (HSCMs), which model networks with degree-heterogeneity \cite{van2018sparse}.

\begin{figure}
\includegraphics[scale=0.25,trim=50 15 0 0]{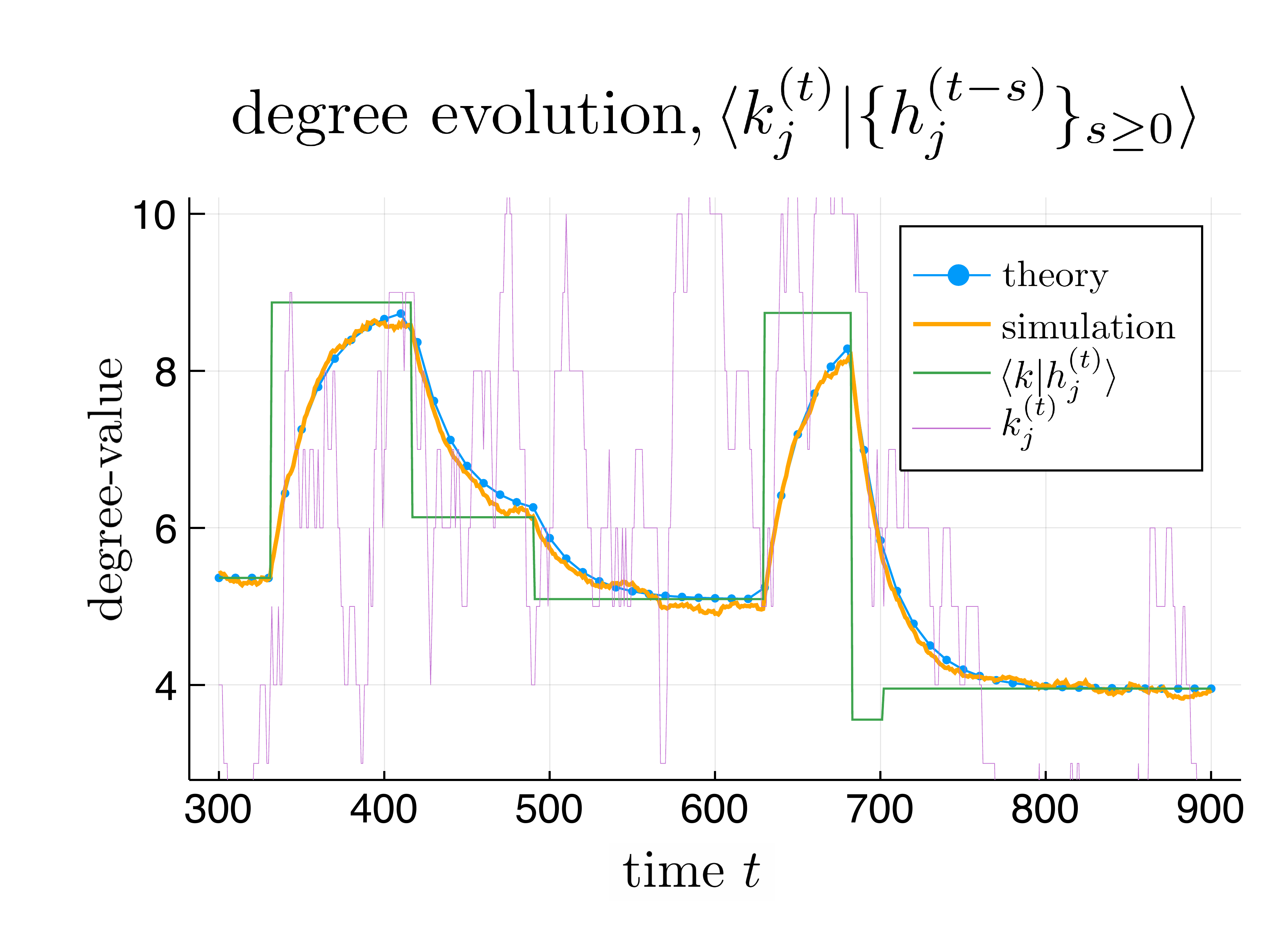}
\caption{{\bf Expected degree over time of a node in a temporal hypersoft 
configuration model with jump-dynamics of hidden variables.} Each node's expected degree ({\bf blue dotted curve}) equilibrates towards its current static-model expected degree ({\bf green solid curve}), as per Equation \ref{eq:degree_dynamics}. In any realization, the {\it actual} degree over time fluctuates ({\bf purple curve}),
but its ensemble-average ({\bf orange solid curve}) behaves as predicted. The average was obtained by simulating $1000$ realizations with $(n,\langle k\rangle,\gamma,\omega,\sigma)=(200,8,2.8,0.04,0.01)$, keeping the HV-trajectory $\{h_j^{(t)}\}_{t=1}^T$ of a single node $j$ fixed across trials.}
\label{fig:degree_dynamics}
\end{figure}

\subsubsection{Static Hypersoft Configuration Model}
\label{sssec:hscm}
The static model we now consider is the {\it hypersoft configuration model} \cite{van2018sparse,voitalov2020weighted} (HSCM), a hyperparametric version of a soft configuration model (SCM). SCMs come in several varieties such as the Chung-Lu model \cite{chung2002connected}, inhomogeneous random graphs \cite{bollobas2007phase}, and the Norros-Reittu model \cite{norros2006conditionally}. Node-pairs connect with $A_{ij}$-values being independent (typically Bernoulli or Poisson distributed), such that on average, each node has a particular degree-value. In hyperparametric SCMs, that degree-value is randomly assigned, according to some specified distribution of expected degrees. For example, one way to obtain SCMs with a degree distribution that is Pareto-mixed Poisson (with, say, power-law tail-exponent $\gamma$ and expected degree $\langle k\rangle$), is for nodes $j\in[n]$ to be assigned hidden variables $h_j\in[h_-,\infty)$ drawn from a Pareto density $\nu(h)=(\gamma-1)h_-^{\gamma-1}h^{-\gamma}$, with minimal HV-value $h_-=(\gamma-2)\langle k\rangle/(\gamma-1)$, and then for node-pairs to be connected with probability
\begin{equation}\label{eq:hscm_affinity}
f(h_i,h_j)=\frac{1}{1+n\langle k\rangle/h_ih_j}\approx \frac{h_ih_j}{n\langle k\rangle},
\end{equation}
the approximation holding when $h_ih_j/n\langle k\rangle\ll 1$. The expected degree of a node $i$ in the static model is
\begin{equation}
\langle k_i|h_i\rangle = (n-1)\int_{h_-}^\infty f(h_i,h)\nu(h)dh \approx h_i.
\end{equation}
The {\it actual} degrees of nodes are sharply peaked around their expected degrees, and thus the above implies that the degree distribution itself likewise has a power-law tail with exponent $\gamma$ and mean $\langle k\rangle$.

\subsubsection{Temporal HSCMs}
Now we consider a temporal version of HSCMs. At each timestep, each node $j$, with probability $\sigma$, resamples its hidden variable $h_j^{(t)}$ from the static-model HV-density $\nu$ (jump-dynamics). Then, each node-pair $ij$ ($1\le i<j\le n$), with probability $\omega$, has its indicator-variable $A_{ij}^{(t)}$ resampled from a Bernoulli of mean $f(h_i^{(t)},h_j^{(t)})$. 

In the static model, the HV-value $h_j$ alone determines the expected degree $\langle k_j|h_j\rangle$. But in the temporal version, the quantity $h_i^{(t)}$ is time-evolving, and the expected degree dynamically trails behind the static-model expected degree, equilibrating at a geometric pace (See Figure \ref{fig:degree_dynamics}):
\begin{equation}\begin{aligned}\label{eq:degree_dynamics}
&\mathbb{E}\left[k_i^{(t)}\left\vert \left\{h_i^{(t-s)}\right\}_{s\ge 0}\right.\right]\\
&=(n-1)\omega\sum_{s\ge 0}(1-\omega)^s\int_{h_-}^\infty f\left(h_i^{(t-s)},h\right)\nu(h)dh\\
&=\omega\sum_{s\ge 0}(1-\omega)^s\left\langle k_i\left\vert h_i^{(t-s)}\right.\right\rangle.\\
\end{aligned}\end{equation}

We can also average the above over all hidden-variable values at timesteps earlier than $t$, to obtain an {\it effective expected degree} that depends only on $h_j^{(t)}$. To do this, we use the probability density of $h_j^{(t-s)}$ given $h_j^{(t)}$ under jump-dynamics:
\begin{equation}
P_s\left(x\left\vert h_j^{(t)}\right.\right)=(1-\sigma)^s\mathbf{1}_{h_j^{(t)}}(x)+\left(1-(1-\sigma)^s\right)\nu(x),
\end{equation}
Averaging Equation \ref{eq:degree_dynamics} over HVs at all timesteps $t-s$ for $s>0$,
\begin{equation}\begin{aligned}
\mathbb{E}\left[k_i^{(t)}\left\vert h_i^{(t)}\right.\right] &= \omega\sum_{s\ge 0}(1-\omega)^s\int_{h_-}^\infty P_s\left(x\left\vert h_i^{(t)}\right.\right)\langle k_i |x\rangle dx\\
&=\alpha_1 \left\langle k_i\left\vert h_i^{(t)}\right.\right\rangle+(1-\alpha_1)\langle k\rangle,\\
\end{aligned}\end{equation}
where $\alpha_1(\sigma,\omega)=\omega\left/\left(1-(1-\omega)(1-\sigma)\right)\right.$. In this case $\alpha_1$ measures the level of equilibration of node-neighborhoods to their expected sizes. Having $\alpha_1\approx 1$ indicates the quasi-static regime whereas $\alpha_1\approx 0$ indicates an averaged-out behavior so that the expected degree of any given node is simply the expected average degree $\langle k\rangle$ of the network.

\subsubsection{Effective connection probabilities in temporal HSCMs}
We now discuss effective connection probabilities in HSCMs. The formula derived in Appendix \ref{app:effective_connection_probabilities} applies, but note that the affinity $f(h,h')$ (Equation \ref{eq:hscm_affinity}) is a function only of the product $\psi=hh'$. Thus we can examine the {\it effective} connection probability as a function of $\psi$, denoted $\bar{f}(\psi)$. In order to calculate $\bar{f}(\psi)$ we first must compute the probability density of a product of hidden variables in past timesteps, given the value of the product at the current timestep. We then sum the expected affinity given the product, weighted by $p_s=\omega(1-\omega)^s$, over all past timesteps $s>0$. These calculations require a variety of intermediate steps, and are described in Appendix \ref{app:product_hidden_variables}.

\subsubsection{Temporal HSCMs discussion}
Note that in HSCMs, non-equilibrium dynamics {\it reduces} degree-heterogeneity; nodes with large HV-values only transiently retain them. Equilibration, on the other hand, allows for a full structural expression of the nodes' internal heterogeneity. This implies that extremely heterogeneous real-world networks, if described by these models, would typically be in the quasi-static regime. We only considered jump-dynamics here (resampling of static-model expected degree-values), but we could alternatively study walk-dynamics, where nodes' HVs undergo Brownian-like motion in a way that preserves $\nu$. This could be achieved straightforwardly as described in \ref{app:simulating_walk_dynamics}, alongside reflecting boundaries as studied in Appendix \ref{app:reflecting}.

\subsection{Temporal Hyperbolic Graphs}
\label{ssec:temporal_hyperbolic_graphs}
In this section we consider a temporal extension of the hyperbolic model \cite{krioukov2010hyperbolic} (the $\mathbb{H}^2$ model, for short), a geometry-based network model simultaneously exhibiting sparsity, clustering, small-worldness \cite{bringmann2016average,friedrich2018diameter}, degree heterogeneity, community structure \cite{faqeeh2018characterizing}, and renormalizability \cite{garcia2018multiscale}. 

\begin{figure*}
\includegraphics[scale=0.5,trim=70 170 0 100]{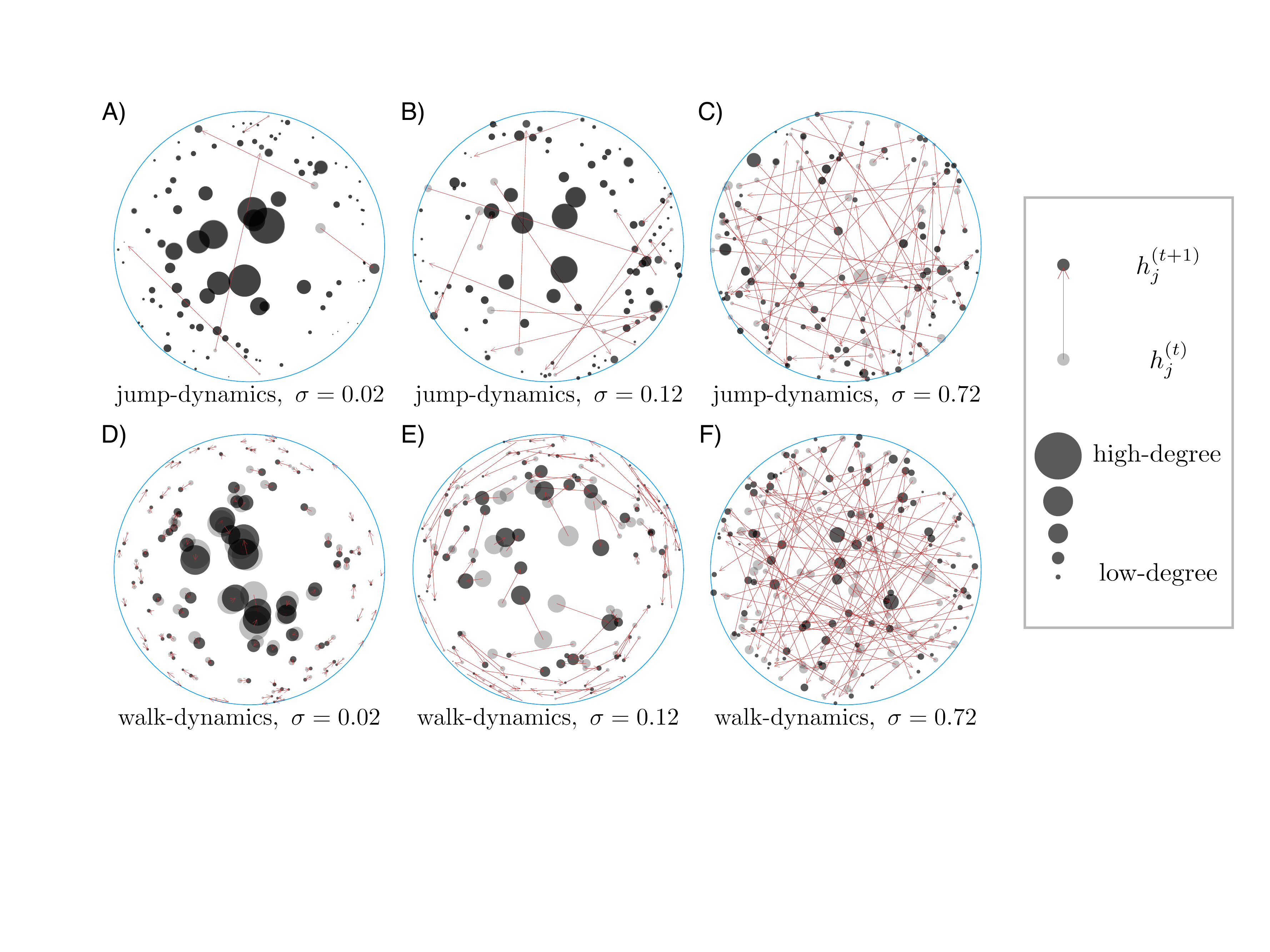}
\caption{{\bf Hidden-variable dynamics of nodes in a temporal $\mathbb{H}^2$ model, at increasing values of $\sigma$, with fixed $\omega=0.1$.} In each subplot, {\bf node-coordinates} for $100$ random nodes are shown at two adjacent timesteps, from a network with parameters $(n,\gamma,\beta,R)=(500,2.2,5,8)$. Each {\bf arrow} points from the coordinate-location of a node at a given timestep ({\bf grey}) to the coordinate-location of the same node at the next timestep ({\bf black}). {\bf Subplots (A,B,C)} depict jump-dynamics (coordinate-resampling with probability $\sigma$, otherwise remaining in place), whereas {\bf subplots (D,E,F)} depict walk-dynamics (all nodes move to neighboring locations, with mean step-length parameterized by $\sigma$). Marker sizes are proportional to node degree-values. For small $\sigma/\omega$ ({\bf subplots A and D}), nodes' existing connections have arisen from approximately the present coordinates, making snapshots closely resemble the static hyperbolic model, as seen e.g. by the exhibited degree-heterogeneity. For larger $\sigma/\omega$ ({\bf subplots B and E}), connections have arisen via mixtures of past and present coordinates, reducing degree-heterogeneity. For very large $\sigma/\omega$ ({\bf subplots C and F}), the system behaves similarly to a temporal Erd\H{o}s-R\'enyi network.}
\label{fig:visH2}
\end{figure*}

\begin{figure}
\includegraphics[scale=0.25,trim=20 0 0 0]{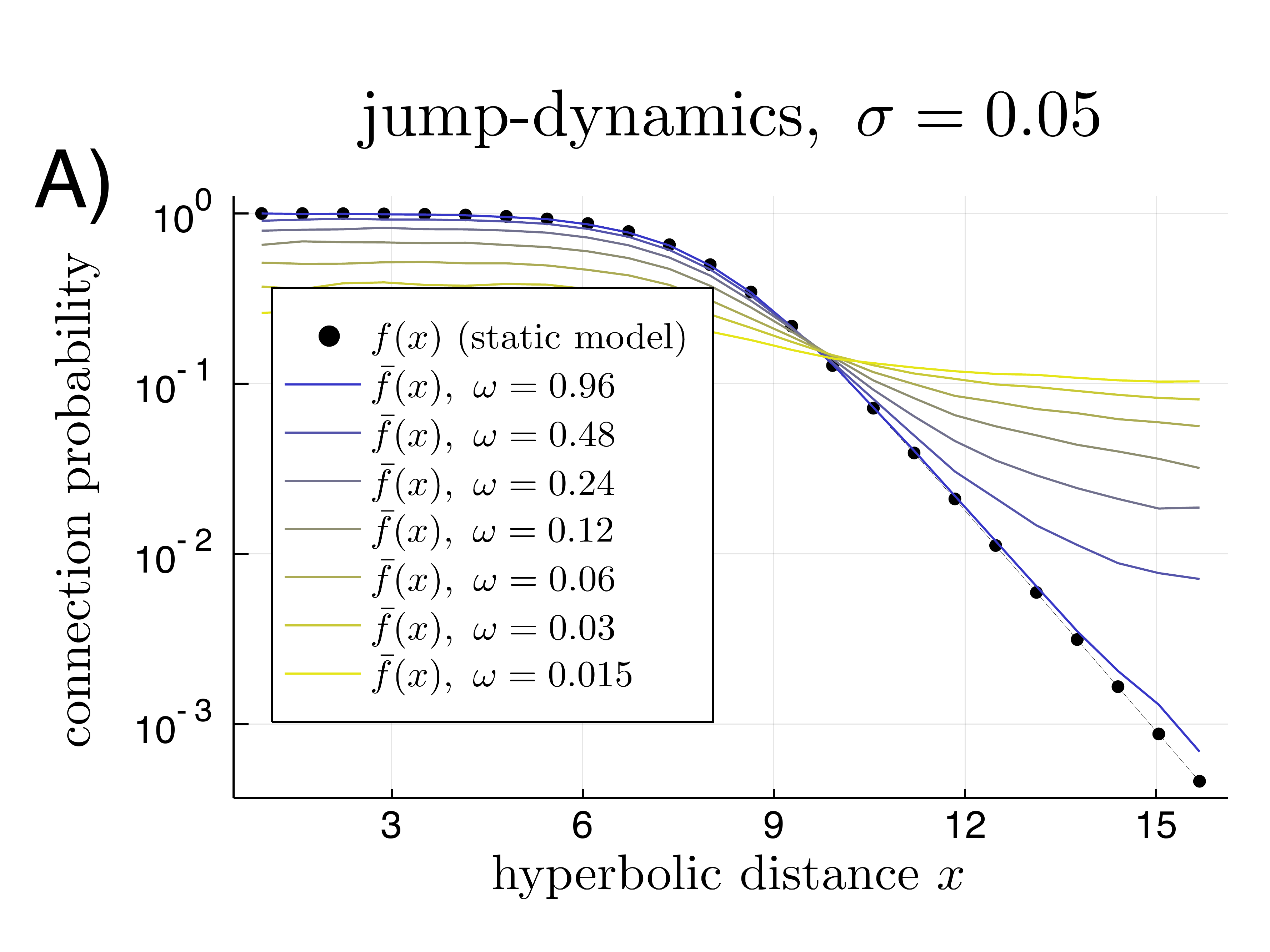}
\includegraphics[scale=0.25,trim=20 0 0 0]{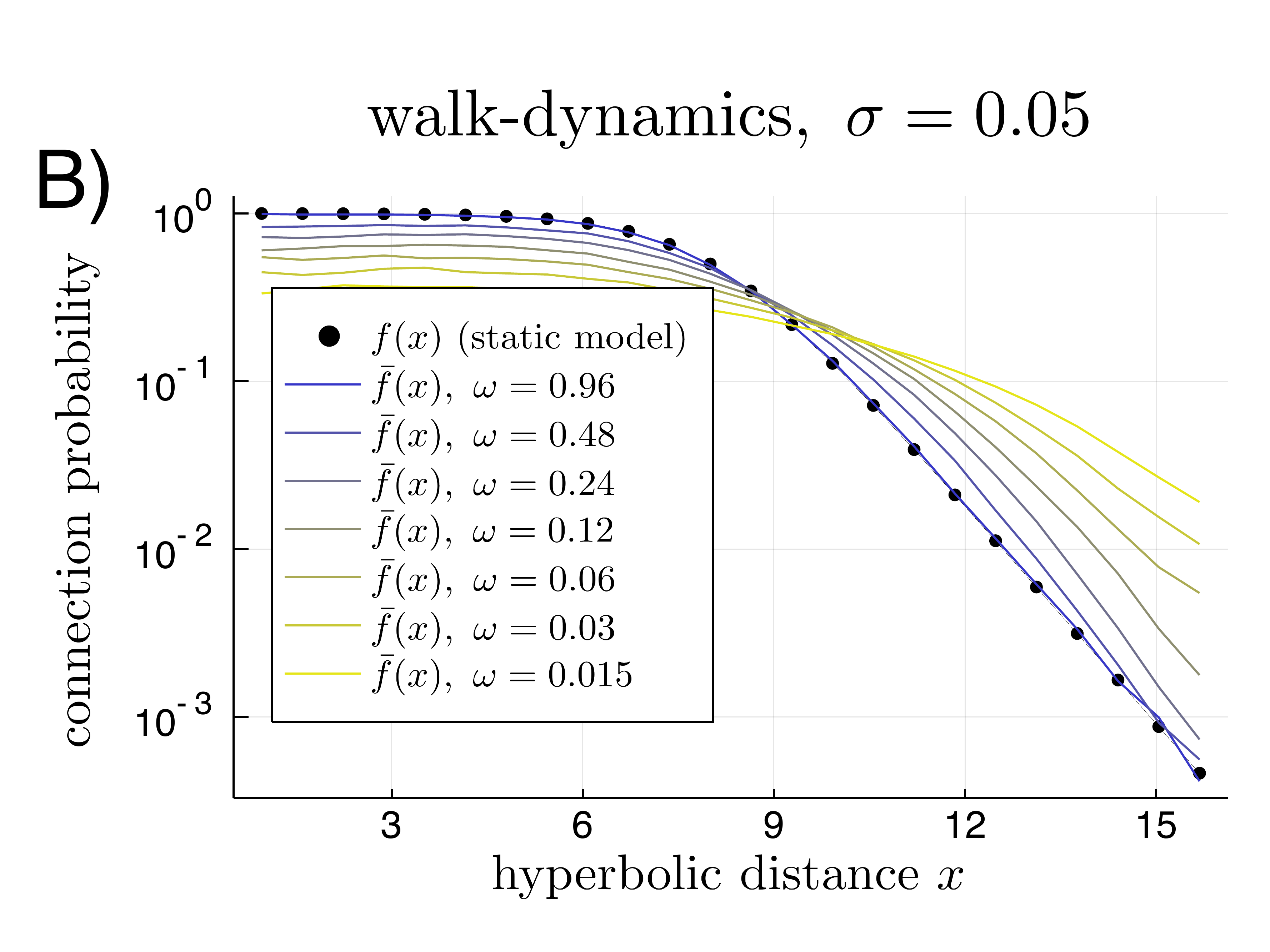}
\caption{{\bf Effective connection probability function $\bar{f}(x)$ in snapshots of a temporal hyperbolic model with $(n,\gamma,\beta,R)=(500,2.2,5,8)$, for various values of $\omega$.} With slower link-resampling (smaller $\omega$), links are increasingly allowed to dynamically stretch before being removed by link-resampling, resulting in deviations from the static-model affinity $f(x)$ ({\bf black dotted line}). {\bf Coloration} of the curve $\bar{f}(x)$ is from yellow to blue as $\omega$ increases. {\bf The upper panel, A)}, shows the case of jump-dynamics of coordinates. {\bf The lower panel, B)}, shows the case of walk-dynamics of coordinates. The choice of coordinate-dynamics is consequential in the non-equilibrium regime, despite each having the same stationary density.}
\label{fig:eff_con_plots}
\end{figure}

\subsubsection{Static $\mathbb{H}^2$ model}
The $\mathbb{H}^2$ model is parameterized by a number of nodes $n$, average degree $\langle k\rangle$, power-law exponent $\gamma$, and inverse-temperature $\beta$ (which tunes the level of clustering). Hidden variables are polar coordinates, $h_j=(\theta_j,r_j)$, namely a radial coordinate $r_j\in[0,R]$ encoding the {\it popularity} of node $j$ and an angular coordinate $\theta_j\in[0,2\pi)$, encoding the {\it similarity} of node $j$ to other nodes. These coordinates are sampled according to separable density $\nu(\theta,r)=\nu_{ang}(\theta)\nu_{rad}(r)$ where angles are distributed uniformly ($\nu_{ang}(\theta)=1/2\pi)$ and radii have an exponentially growing density, 
\begin{equation}
\nu_{rad}(r)=\frac{\gamma-1}{2}\frac{\sinh\left(\frac{\gamma-1}{2}r\right)}{\cosh\left(\frac{\gamma-1}{2}R\right)-1}, 
\end{equation}
where $R=R(n,\langle k\rangle,\beta,\gamma)$ is selected so that the mean degree is $\langle k\rangle$. The static-model affinity of node-pair $ij$ is a Fermi-Dirac function \cite{kardar2007statistical} (a sigmoid) of the hyperbolic geodesic distance $x_{ij}$ between $i$ and $j$,
\begin{equation}\label{eq:hyp_con_pro}
f(h_i,h_j)=f(x_{ij})=1\left/
\left(1+e^{(\beta/2)\left(x_{ij}-R\right)}\right)\right.,
\end{equation}
where $x_{ij}=x_{ij}(h_i,h_j)$ is given by
\begin{equation}\begin{aligned}
\cosh (x_{ij})=&\cosh(r_i)\cosh(r_j)\\
&-\sinh(r_i)\sinh(r_j)\cos(\theta_{ij}),
\end{aligned}\end{equation}
with $\theta_{ij}=\pi-\left\vert\pi-|\theta_i-\theta_j|\right\vert$. The connection probability and coordinate-density in this model result in power-law degree distributions (but could also give rise to other degree distributions if the radial coordinate-density was different), a similar feature to that exhibited by HSCMs -- but also, the geometry arising from inclusion of the angular coordinate yields a large clustering coefficient and spatially localized link-structure, making this model also similar to standard RGGs. Increasing the parameter $\beta$ yields more localized link-structure, approaching a step function as $\beta\rightarrow\infty$, leaving in that case an RGG (see Section \ref{sssec:rgg}) on the hyperbolic disk. As $\beta\rightarrow 0$, typical link-lengths approach the system size and the model behaves similarly to the HSCM (see Section \ref{sssec:hscm}).

\subsubsection{Temporal $\mathbb{H}^2$ model}
To temporally extend the $\mathbb{H}^2$ model, we allow coordinate dynamics so that each node $j$ exhibits a trajectory in the hyperbolic disk, $h_j^{(t)}=(\theta_j^{(t)},r_j^{(t)})$ for $t\in[T]$. For jump-dynamics, each node jumps to a random location according to density $\nu(\theta,r)$, with probability $\sigma$ each timestep. For walk-dynamics, each node $j$ steps to a random location $h_j^{(t+1)}$ having angular and radial coordinates adjusted to relatively closeby values, with increasingly large steps for larger $\sigma$-values; we describe the details of $\mathbb{H}^2$ walk-dynamics in Appendix \ref{app:H2_walk_dynamics}. Dynamics of nodes on the hyperbolic disk is visualized in Figure \ref{fig:visH2}, for both jump-dynamics and walk-dynamics. For $\sigma\ll 1$, nodes rarely resample their coordinates (in jump-dynamics) and step to only very localized regions (in walk-dynamics). On the other hand for $\sigma\approx 1$, almost all nodes resample their coordinates at each timestep (in jump-dynamics) or move to a nearly-randomized location (in walk-dynamics). We note that many other natural and interesting choices for HV-dynamics exist, as we discuss in Section \ref{sec:discussion} and Appendix \ref{app:H2_walk_dynamics}. 

\subsubsection{Effective connection probabilities in the temporal $\mathbb{H}^2$ model}
In the temporal $\mathbb{H}^2$ model considered here, the effective connection probability $\bar{f}(x)$ no longer remains in the standard Fermi-Dirac form of $f(x)$ (see Figure \ref{fig:eff_con_plots}). With decreasing $\omega/\sigma$, the connection probability function smooths out and extends to a longer range due to links being stretched more rapidly (for walk-dynamics), or more frequently (for jump-dynamics). This effect is more uniform and extends all the way out to long ranges for jump-dynamics, whereas it is more localized for walk-dynamics, for any given non-equilibrium value of $(\sigma,\omega)$.

Since the coordinates of $\mathbb{H}^2$ reflect popularity and similarity attributes, the effective connection probability and other non-equilibrium effects arising when outside of the quasi-static regime have specific interpretations. The set of current links arose from nodes having been connected at past timesteps when their previous similarity attributes were compatible (small hyperbolic distance); in real networks, such links may persist into the future even if the similarity attributes change. For instance with social networks, consider friendships on Facebook, followers on Twitter, or author collaborations: similarity between connected pairs may decrease over time, but they tend to remain connected. Likewise, it could take some time for two people that become more similar to discover one another and to connect, in an online or traditional social network.

\subsubsection{Temporal $\mathbb{H}^2$ model discussion}
Outside of the quasi-static regime, snapshots $G^{(t)}$ do not fully resemble the static $\mathbb{H}^2$ model -- the Equilibrium Property is in general violated (despite the fact that {\it each link} was formed via the static-model connection probability corresponding to the pairwise distance at the time of that link's formation). This phenomenon results in reduced clustering because links become spread out across the space rather than being localized amongst neighboring groups of nodes. Degree-heterogeneity is also suppressed, as is the case for the temporal HSCM (see Section \ref{ssec:temporal_hypersoft_configuration}), because nodes accumulating large numbers of links due to being near the disk's center do not stay near the disk's center indefinitely. Clustering and heterogeneity arise in the {\it static} $\mathbb{H}^2$ model due to the correlations in links from the underlying geometry. But in the static model, all links (and non-links) arise from the {\it same} underlying coordinate-configuration. When coordinates are dynamical, these correlations are weaker; nodes are linked with probabilities arising as a mixture of past and present coordinate-configurations.

\section{Link-updating in response to hidden-variable dynamics}\label{sec:link_response}
Finally, we describe an additional dynamical mechanism that can be incorporated to achieve the Equilibrium Property {\it exactly} in temporal hidden-variables models, while retaining the Persistence Property, for all values of $\sigma$ and $\omega$: links are updated {\it directly in response} to changes in hidden variables, rather than only through link-resampling, to keep connection probabilities up-to-date (we refer to this mechanism as link-response). In this model variant, $G^{(t+1)}$'s probability distribution depends on each of $G^{(t)}$, $H^{(t+1)}$, and $H^{(t)}$, rather than on just the former two. We illustrate the mechanism at first in the case of $\omega=0$. Suppose node-pair $ij$ has a link with probability $p_{ij}=f(h_i,h_j)$, and that HVs $(h_i,h_j)$ are updated to become $(h'_i, h'_j)$ in the next timestep. To ensure that the pair is then connected with probability $p_{ij}'=f(h'_i,h'_j)$, we selectively delete now-less-likely edges between connected pairs and selectively add now-more-likely edges between unconnected pairs. In particular: 
\begin{itemize}
\item[a)] If $p_{ij}'\ge p_{ij}$, then $A_{ij}=1\Rightarrow A'_{ij}=1$, and $A_{ij}=0\Rightarrow$ add link with probability $q^+_{ij}$,
\item[b)] If $p_{ij}'\le p_{ij}$, then $A_{ij}=0\Rightarrow A'_{ij}=0$, and $A_{ij}=1\Rightarrow$ remove link with probability $q^{-}_{ij}$.
 \end{itemize}
The outcome needs to result in $\mathbb{P}(A'_{ij}=1|h_i',h_j')=p'_{ij}$. Thus,
\begin{itemize}
\item[a)] If $p_{ij}'\ge p_{ij}$, the new connection probability satisfies $p'_{ij}=p_{ij}+(1-p_{ij})q^+_{ij}$. Hence, $q^+_{ij}=1-\frac{1-p'_{ij}}{1-p_{ij}}$.
\item[b)] If $p_{ij}'\le p_{ij}$, the new connection probability satisfies $1-p'_{ij}=p_{ij}q^{-}_{ij}+(1-p_{ij})$. Hence, $q^-_{ij}=1-\frac{p'_{ij}}{p_{ij}}$.
\end{itemize}
Note that if $p'_{ij}=p_{ij}$, then $q^{+}_{ij}=q^{-}_{ij}=0$; no links will form or break unless pairwise affinities change. Denoting $p_{ij}^{(t)}=f(h_i^{(t)},h_j^{(t)})$ for $t\in\{1,...,T\}$, the graph transition probability given $\mathbf{H}$ becomes:

\begin{equation}
\mathcal{P}_G\left(\left. G^{(t+1)}\right\vert G^{(t)},\mathbf{H}\right) = \prod_{1\le i<j\le n}Y_{ij}\left(A_{ij}^{(t+1)}\left\vert A_{ij}^{(t)},\mathbf{H}\right)\right.,
\end{equation}
with $Y_{ij}:\{0,1\}\rightarrow[0,1]$ denoting the conditional adjacency-element probability distribution. For any $\omega\in[0,1]$, we have:
\begin{equation}
Y_{ij}\left(1\left\vert A_{ij}^{(t)},\mathbf{H}\right)\right.=\omega p_{ij}^{(t+1)}+(1-\omega)K_{ij}\left(A_{ij}^{(t)},\mathbf{H}\right),
\end{equation}
where $K_{ij}(A_{ij}^{(t)},\mathbf{H})$ incorporates the link-response dynamics:
\begin{equation}
\begin{aligned}
K_{ij}\left(A_{ij}^{(t)},\mathbf{H}\right)=&\mathbf{1}\left\{p_{ij}^{(t+1)}\ge p_{ij}^{(t)}\right\}\left(q^+_{ij}\left(1-A_{ij}^{(t)}\right)+A_{ij}^{(t)}\right)\\
&+\mathbf{1}\left\{p_{ij}^{(t+1)}\le p_{ij}^{(t)}\right\}(1-q^{-}_{ij})A_{ij}^{(t)}.
\end{aligned}
\end{equation}

With the inclusion of link-response, arbitrary static hidden-variable networks can be extended to temporal settings while satisfying the Equilibrium Property exactly (See Appendix \ref{sec:link_response} for a full derivation), and the Persistence Property in a tunable fashion. Allowing $\omega>0$ does not alter the Equilibrium Property's exact validity, and it provides a more tunable level of structural persistence. 

With $G^{(t)}$ indistinguishable from a static-model realization, all non-equilibrium phenomena of the types discussed in \ref{ssec:temporal_stochastic_block}, \ref{ssec:temporal_random_geometric}, \ref{ssec:temporal_hypersoft_configuration}, and \ref{ssec:temporal_hyperbolic_graphs} are prevented -- this can either enhance or hinder Qualitative Realism, depending on the context. If a single node's HV is changed, it will need to re-evaluate connections to {\it all} other nodes for which affinities have changed. This could be realistic in some cases, since nodes themselves may be at the most liberty to re-evaluate their connections. In other cases, more gradual structural transitions may be preferred. This model-variant could thus serve well as a temporal null model, especially for temporal networks with snapshots well-described by an SHVM. Despite {\it structure} of THVMs with link-response being identical to that of SHVMs, all {\it dynamical} features are open for study and for comparison to real-world networks.

\section{Related Work}\label{sec:related_work}
We briefly review existing lines of research related to our study.

Several temporal network models are worth mentioning. Temporal analogs of specific static models have been considered \cite{zhang2017random,mandjes2019dynamic,peixoto2017modelling,xu2014dynamic,xu2015stochastic,matias2017statistical,pensky2019spectral,ghasemian2016detectability,barucca2018disentangling}, many of which preserve the Equilibrium Property. Most such models have non-dynamic node properties, yielding models related to edge-Markovian networks \cite{clementi2009information,clementi2010flooding,whitbeck2011performance,du2016continuous,lamprou2018cover} and dynamic percolation \cite{steif2009survey,khoshnevisan2008dynamical,peres1998number}. The dynamic-$\mathbb{S}^1$ model \cite{papadopoulos2019latent} is a temporal extension of the static $\mathbb{S}^1$ model \cite{krioukov2010hyperbolic} consisting of a sequence of independent samples with HVs partially inferred from real data and partially synthetically generated; the dynamics therein resembles THVMs with $\omega=1$ and $\sigma=0$, but with varying average degree parameter across snapshots. Although it is common practice to extend static-model concepts to temporal settings \cite{nicosia2013graph,ortiz2017navigability,taylor2017eigenvector,kim2012temporal,pan2011path,li2017fundamental,liu2014controlling,perra2012random,paranjape2017motifs,holme2016temporal,liu2018epidemic,nadini2018epidemic,masuda2017temporal,sun2015contrasting,li2018opinion,dunlavy2011temporal,dhote2013survey,sarzynska2016null,gauvin2014detecting}, many models of temporal networks are instead derived from first principles \cite{zino2016continuous,pozzana2017epidemic,da2015slow,alessandretti2017random,rizzo2016innovation,starnini2014temporal}, and focus primarily on inference techniques, real-world applicability \cite{silvescu2001temporal,lebre2010statistical,hanneke2010discrete,mellor2019event}, and/or the effects of temporality on spreading \cite{perra2012activity,clementi2010flooding}.

Most relevant to THVMs are several existing works with dynamic HVs that influence link-dynamics. Several dynamic latent space models \cite{sewell2016latent,kim2018review,sarkar2007latent,sarkar2006dynamic} exist, as do dynamic random geometric graphs \cite{peres2013mobile} (the latter being continuous-time and infinite-space, with nodes sprinkled as a Poisson process \cite{miles1970homogeneous,moller2007modern,reitzner2013poisson} and undergoing Brownian motion \cite{varadhan2007stochastic}, with links remaining up-to-date as for THVMs with $\omega=1$). A model with both dynamic HVs and persistent links \cite{mazzarisi2020dynamic} was recently introduced, alongside rigorous inference techniques and applications -- but not in reference to static network models. Other studies investigated spreading on dynamic RGG-like graphs \cite{buscarino2008disease,clementi2015parsimonious}. A few versions of dynamic SBMs are of particular relevance; in one such paper \cite{ghasemian2016detectability}, the model is a case of the temporal hyper-SBM studied in Section \ref{ssec:temporal_stochastic_block} with complete edge-resampling ($\omega=1$). Another study was of a temporal hyper-SBM with $\omega<1$ which thus exhibits both link-persistence and group-assignment-persistence \cite{barucca2018disentangling}, influencing performance of community detection algorithms and motivating the development of new ones. Another area of relevant work is the rapidly emerging area of {\it dynamic graph embeddings} \cite{bian2019network,goyal2018dynamicgem,lu2019temporal,xie2020survey,haddad2019temporalnode2vec,jin2020static,spasov2020grade, chen2019dynamic,cheng2020dynamic,kim2018review,zhu2016scalable,lee2020dynamic,singer2019node,kumar2018learning}, related to the task of {\it inference of hidden-variable trajectories} \cite{papadopoulos2014network}. 

We also note some additional works that are less-directly related to ours. Network-rewiring and MCMC algorithms are widely used to sample static networks \cite{van2010influence,young2017construction,bannink2019switch,bhamidi2008mixing,demuse2019mixing}; in stationarity, these can be viewed as temporal networks satisfying the Equilibrium Property, with a level of persistence tunable via the number of iterations between adjacent snapshots. Adaptive network models (for instance, SIS-dynamics \cite{pastor2015epidemic} alongside contact-switching \cite{risau2009contact,piankoranee2018effects}), have dynamic node-properties that evolve with time and guide network evolution, a commonality with THVMs. Networks with node-growth and node-removal  \cite{dorogovtsev2002evolution,moore2006exact,bauke2011topological,becchetti2020expansion} have dynamic node-properties (degree-values as opposed to hidden variables) that influence link-formation. In the fitness model of growing networks \cite{bianconi2001competition}, static HVs and dynamic degrees both govern connection probabilities. Some static network models admit dual growing formulations \cite{krioukov2013duality} -- analogously, if the Equlibrium Property holds, THVM snapshots can be seen as dynamically produced static-model samples.

\section{Discussion}\label{sec:discussion}
In this work we have studied temporal network models that are natural counterparts of static hidden-variables models, obtained by inclusion of a dynamic mechanism for node-characteristics (jump-dynamics or walk-dynamics) and dynamic mechanism for link-structure (link-resampling). Due to the wide generality of the static hidden-variables framework, many popular static network models can be made temporal as THVMs.

With a single source of randomness in the static model, which includes $\omega=1$ with deterministic connectivity (Section \ref{ssec:complete_edge_renewal}) and $\sigma=0$ with fixed initial HVs (Section \ref{ssec:edge_independent}), the Equilibrium Property is exactly satisfied and the Persistence Property is controllable. If, however, the static model has two layers of randomness and links are not completely refreshed each timestep ($\sigma>0$ and $\omega<1$), THVM snapshots are {\it not} in general distributed according to the static model. Rather, numerous structural deviations arise, due to links falling out-of-equilibrium with respect to hidden variables -- for instance, the effective connection probability $\bar{f}(h,h')$ can substantially differ from the affinity function $f(h,h')$ (see Figures \ref{fig:RGG_effective} and \ref{fig:eff_con_plots}). Despite violating the Equilibrium Property, such models arise naturally and exhibit Qualitative Realism in interesting ways  -- for instance, the appearance of long-ranged links in temporal RGGs (Section \ref{sssec:tsbm_effective}) and inter-group links in temporal hyper-SBMs (Section \ref{sssec:tsbm_effective}). An exception to the non-equilibrium dynamics arises in the quasi-static regime (Section \ref{ssec:quasistatic}) in which case the Equilibrium Property is {\it approximately} satisfied, due to all $A_{ij}^{(t)}$-values arising from an HV-configuration closely resembling $H^{(t)}$. A second exception arises if we add a third dynamical mechanism (Section \ref{sec:link_response}), namely link-updating in {\it direct} response to HV-changes, which allows {\it exact} satisfaction of the Equilibrium Property (see Appendix \ref{app:stationarity_with_link_response}) for all $(\sigma,\omega)$. Both situations also lend themselves to tunable satisfaction of the Persistence Property, governed $\sigma$ and $\omega$. 

An assortment of possible modifications, improvements, and extensions are worth mentioning. Although many questions are open within present framework, altered dynamics could also be considered. For HV-dynamics, correlated motion akin to Langevin dynamics \cite{schlick2010molecular,flores2018similarity} could provide insight into the formation and persistence of communities. Altered link-structure and link-dynamics could be considered as well: some examples include directed and/or weighted links, node-centric link-resampling dynamics \cite{jacob2017contact}, or pairwise-individualized resampling rates. Continuous-time formulations of THVMs could allow some theoretical simplifications; continuous time is used in studies of dynamical percolation \cite{steif2009survey,olle1997dynamical,garban2018scaling} and edge-Markovian networks \cite{clementi2010flooding,whitbeck2011performance,de2013relevance,roberts2018exceptional,rossignol2020scaling}, which could each be extended to a THVM-like framework by introducing hidden variables. Our results can also inform future studies of adaptive networks \cite{bassler2019coevolution,sayama2013modeling,choromanski2013scale,caldarelli2008self,papadopoulos2017development}; THVMs provide a simple setting in which dynamic node-properties influence network-evolution. Understanding such settings will provide a baseline for what to expect when coevolutionary feedbacks are also present. An example of real-world links influencing node-properties is social influence, whereby acquainted pairs can become more similar over time \cite{leenders1997longitudinal,eom2016concurrent} -- or geographically move to closer-by coordinate locations. The inclusion of interdependencies relating to dynamical processes \cite{mancastroppa2020active,ichinose2018reduced} can allow for more interesting dynamics and realism, but at the cost of increased model complexity. 

Real-world networks have dynamic node-properties that influence dynamics of link-structure. Examples of such phenomena were set forth in Section \ref{sec:intro}, ranging across a wide variety of systems and scales. One direct real-world application of THVMs could be to serve as null models \cite{gotelli2001research,sarzynska2016null} for evolving networks with dynamic node-properties \cite{kim2018review}. Dynamic embedding methods \cite{bian2019network,goyal2018dynamicgem,lu2019temporal,xie2020survey,haddad2019temporalnode2vec,jin2020static,spasov2020grade, chen2019dynamic,cheng2020dynamic,zhu2016scalable,lee2020dynamic,singer2019node,kumar2018learning}, or generalizations of inference methods from dynamic SBMs \cite{barucca2018disentangling}, could potentially allow retrieval of $\mathbf{H}$ (and perhaps also $\sigma$, $\omega$, and $f$) from an observed $\mathbf{G}$. Links of real evolving networks may not in general be fully equilibrated relative to the current set of node-characteristics, which is a dynamical behavior exhibited by THVMs {\it outside of the quasi-static regime}. Hence in some cases, the Equilibrium Property and Qualitative Realism may be {\it in conflict}, implying that caution should be used when applying static models to snapshots of evolving networks. That said, static models do in many cases accurately describe such snapshots; the internet, for example, has exhibited a clear power-law degree-tail for decades \cite{papadopoulos2014network,siganos2003power}, evidently remaining in equilibrium from the perspective of THVMs (see the discussion in \ref{ssec:temporal_hypersoft_configuration}). 

Overall, we expect that the present study will usefully inform general classifications of real-world networks according to the dynamics of node-properties and of how those properties influence link-dynamics.

\section{Acknowledgements}
We thank B. Klein, S. Redner, M. Shrestha, L. Torres, R. Van der Hofstad, and I. Voitalov for useful discussions and suggestions. This work was supported by ARO Grant Nos. W911NF-16-1-0391 and W911NF-17-1-0491, and by NSF Grant Nos. IIS- 1741355 and DMS-1800738. F.P. acknowledges support by the TV-HGGs project (OPPORTUNITY/0916/ERC-CoG/0003), funded through the Cyprus Research and Innovation Foundation.

\appendix

\section{Effective connection probabilities}
\label{app:effective_connection_probabilities}
Here we calculate effective connection probabilities for general THVMs with HVs evolving by jump-dynamics (HV-resampling with probability $\sigma$). We define the effective connection probability $\bar{f}(h,h')$ to be the probability of $A_{ij}^{(t)}=1$ given $h_i^{(t)}=h$ and $h_j^{(t)}=h'$, in the limit as $t\rightarrow\infty$. That is,
\begin{equation}
\bar{f}(h,h')=\lim_{t\rightarrow\infty}\mathbb{P}\left(A_{ij}^{(t)}=1\left\vert h_i^{(t)}=h,h_j^{(t)}=h'\right.\right),
\end{equation}
where the limit $t\rightarrow\infty$ is to wash out any initial condition. Due to the edge-resampling dynamics, the current value of $A_{ij}^{(t)}$ arose from being last resampled at some time $t-s$, with $s$ being a random nonnegative integer having distribution $p_s=\omega (1-\omega)^s$ (where $\omega$ is the probability of link-resampling at any given timestep). The effective connection probability is given by
\begin{equation}\label{eq:eff_con_pro_exp}
\bar{f}(h,h')=\sum_{s\ge 0} p_s \mathbb{E}\left[f\left(h_i^{(t-s)},h_j^{(t-s)}\right)\left\vert h_i^{(t)}=h,h_j^{(t)}=h'\right.\right].
\end{equation}
To evaluate the above, we introduce a density $P_s(x|h)$, namely the density of $h_{i}^{(t-s)}$ (evaluated at $x$) given $h_i^{(t)}=h$. In our case, by jump-dynamics and conditioning on $h_i^{(t)}=h$, we have
\begin{equation}\label{eq:past_HV_density}
P_s(x|h) =(1-\sigma)^s\mathbf{1}_{h}(x)+\left(1-(1-\sigma)^s\right)\nu(x),
\end{equation}
because $h$ will have arisen from $x$ after $s$ timesteps via either (a) zero jumps having occurred, that event having probability $(1-\sigma)^s$, or via (b) {\it at least one} jump having occurred, in which case the density is completely randomized to $\nu(x)$. The expectation value appearing in Equation \ref{eq:eff_con_pro_exp} is equal to
\begin{equation}
\begin{aligned}
& \mathbb{E}\left[f\left(h_i^{(t-s)},h_j^{(t-s)}\right)\left\vert h_i^{(t)}=h,h_j^{(t)}=h'\right.\right] \\
 &= \int_{\mathcal{X}}\int_{\mathcal{X}} f(x,x')P_s(x|h)P_s(x'|h')dxdx',
 \end{aligned}
\end{equation}
which, using Equation \ref{eq:past_HV_density} and integrating over $(x,x')$, evaluates to:
\begin{equation}\label{eq:result_of_integration}
\begin{aligned}
& (1-\sigma)^{2s}f(h,h')\\
&+(1-\sigma)^s(1-(1-\sigma)^s)\left(\langle f(\cdot ,h')\rangle+\langle f(h,\cdot)\rangle\right)\\
&+(1-(1-\sigma)^s)^2\langle f\rangle,\\
\end{aligned}\end{equation}
where $\langle f(\cdot,h)\rangle=\langle f(h,\cdot)\rangle=\int_{\mathcal{X}}f(h,x)\nu(x) dx$ and $\langle f\rangle=\int_{\mathcal{X}^2}f(x,x')\nu(x)\nu(x')dxdx'$. Finally, plugging Equation \ref{eq:result_of_integration} back into Equation \ref{eq:eff_con_pro_exp}, using $p_s=\omega(1-\omega)^s$ and summing the geometric series that appear ($\sum_{s\ge 0}y^s=1/(1-y)$), we obtain
\begin{equation}\label{eq:eff_con_pro_general}
\begin{aligned}
\bar{f}(h,h')&=\alpha_2 f(h,h')\\
&+(\alpha_1-\alpha_2)\left(\langle f(\cdot ,h')\rangle+\langle f(h,\cdot)\rangle\right)\\
&+(1-2\alpha_1+\alpha_2)\langle f\rangle,
\end{aligned}
\end{equation}
where $\alpha_b=\alpha_b(\sigma,\omega)$ for $b\in\{1,2\}$ are given by
\begin{equation}\label{eq:alpha_beta}
\alpha_b(\sigma,\omega) = \frac{\omega}{1-(1-\omega)(1-\sigma)^b}.
\end{equation}
As an aside, we note that the average degree of the network is independent of $(\sigma,\omega)$. This can be seen by averaging Equation \ref{eq:eff_con_pro_exp} over $h$ and $h'$ and making use of $\int_{\mathcal{X}}P_s(x|h)\nu(h)dh=\nu(x)$ (which is true because $P_s(x|h)$ describes the stationary distribution, regardless of whether we consider walk-dynamics or jump-dynamics). The result is $\langle f\rangle$, regardless of $\sigma$ and $\omega$. This can be seen more directly in the case of jump-dynamics by averaging Equation \ref{eq:eff_con_pro_general} over $h$ and $h'$. 

\section{Temporal RGG effective connection probability}
\label{app:temporal_rgg_effective}
This section contains calculations of the effective connection probability for random geometric graphs on the unit interval with periodic boundaries and jump-dynamics. This result could be obtained from Equation \ref{eq:eff_con_pro_general}, but we show here an alternate derivation. The effective connection probability as a function of distances is defined as the probability of two nodes {\it being connected} given that they are a distance $\mathrm{d}_{ij}^{(t)}=x$ apart, as $t\rightarrow\infty$:
\begin{equation}
\bar{f}(x)=\lim_{t\rightarrow\infty} \mathbb{P}\left(\left. A_{ij}^{(t)}=1\right\vert\mathrm{d}_{ij}^{(t)}=x\right).
\end{equation}
To calculate the above, we introduce the probability density on distances between node-pairs $s$ timesteps prior to when the distance-value is $x$, denoted $P_s(y|x)$. We make use of the fact that $\mathrm{d}_{ij}^{(t)}$ can evolve in either of two ways: with probability $(1-\sigma)^2$ each timestep, {\it neither} $i$ {\it nor} $j$ jumps, and thus the density is preserved. Otherwise, one or both do jump, and their distance becomes completely randomized. The stationary density of distance $x$ is the uniform on $[0,1/2]$, {\it i.e.}, equal to $2$ for all $x\in[0,1/2]$. In a single time-advancement, jump-dynamics thus yields
\begin{equation}
P_1(y|x)=(1-\sigma)^{2}\mathbf{1}_{x}(y)+2\left(1-(1-\sigma)^{2}\right).
\end{equation}
Iterating the above logic, $P_s(y|x)$ has two contributions: either neither node jumps at any time, or at least one node jumps at least once. Therefore,
\begin{equation}
P_s(y|x)=(1-\sigma)^{2s}\mathbf{1}_{x}(y)+2\left(1-(1-\sigma)^{2s}\right).
\end{equation}
We can compute $\bar{f}(x)$ via averaging the affinity $f(h_i,h_j)=\mathbf{1}\{\mathrm{d}_{ij}^{(t)}\le r\}$ over the distance-variable. That is,
\begin{equation}
\bar{f}(x)=\sum_{s\ge 0}p_s\mathbb{E}\left[\mathbf{1}\left\{\mathrm{d}_{ij}^{(t-s)}\le r\right\}\left\vert \mathrm{d}_{ij}^{(t)}=x\right.\right],
\end{equation}
where the expectation term is
\begin{equation}\begin{aligned}
&\mathbb{E}\left[\mathbf{1}\left\{\mathrm{d}_{ij}^{(t-s)}\le r\right\}\left\vert \mathrm{d}_{ij}^{(t)}=x\right.\right] \\
&= \int_{0}^{1/2} P_s(y|x)\mathbf{1}\{y\le r\}dy\\
&=\int_0^{r} \left((1-\sigma)^{2s}\mathbf{1}_{x}(y)+2\left(1-(1-\sigma)^{2s}\right)\right)dy\\
&=(1-\sigma)^{2s}\mathbf{1}\{x\le r\}+2r\left(1-(1-\sigma)^{2s}\right).
\end{aligned}\end{equation}
Let $s\in\{0,1,...\}$ be the delay since any given edge-indicator was last resampled. Recall that $s$ has distribution $p_s=\omega(1-\omega)^{s}$. Then, using the above, we find that the effective connection probability for 1D RGGs with jump-dynamics is
\begin{equation}\begin{aligned}
\bar{f}(x)&=\omega\sum_{s\ge 0}(1-\omega)^s(1-\sigma)^{2s}\mathbf{1}\{x\le r\}\\
&+\omega\sum_{s\ge 0}(1-\omega)^s2r\left(1-(1-\sigma)^{2s}\right)\\
&=\alpha_2 \mathbf{1}\{x\le r\}+(1-\alpha_2)2r,
\end{aligned}\end{equation}
with $\alpha_2=\alpha_2(\sigma,\omega)$ arising from having evaluated sums of geometric series of the form $\sum_{s\ge 0} ((1-\omega)(1-\sigma)^2)^s$:
\begin{equation}
\alpha_2(\sigma,\omega)=\frac{\omega}{1-(1-\omega)(1-\sigma)^2}.
\end{equation}

\section{Effective connection probability in terms of products of hidden variables}
\label{app:product_hidden_variables}
This section describes effective connection probabilities arising in temporal HSCMs, as studied in Section \ref{ssec:temporal_hypersoft_configuration}. The static-model affinity $f$ is a function of the {\it product} of hidden variables, motivating study of the effective connection probability $\bar{f}$ as a function of the product of HVs as well.

Consider one-dimensional hidden variables $\{h_j\}_{j\in[n]}$ each distributed uniformly on $\mathcal{X}=[0,1]$. This is applicable to HSCMs via the CDF-transform of arbitrary 1D probability densities: if $h$ has density $\nu$, then $u=F(h)=\int_{h_-}^h \nu(h')dh'$ is distributed uniformly on $[0,1]$ ($h_-$ is the minimum value of $h$). Denote $P_s(\phi|\psi)$ as the probability density of $\phi=h_i^{(t-s)}h_j^{(t-s)}$ for some arbitrary pair $ij$ given that $h_i^{(t)}h_j^{(t)}=\psi$. Then,
\begin{equation}
\bar{f}(\psi)=\omega \sum_{s\ge 0}(1-\omega)^s\int_{0}^1P_s(\phi|\psi)f(\phi)d\phi.
\end{equation}
For products of HVs each independently undergoing jump-dynamics, we have
\begin{equation}\begin{aligned}
P_s(\phi|\psi)&=(1-\sigma)^{2s}\mathbf{1}_{\psi}(\phi)\\
&+\left(1-(1-\sigma)^s\right)(1-\sigma)^sp_{1}(\phi|\psi)\\
&+\left(1-(1-\sigma)^s\right)^2\mu(\phi),
\end{aligned}\end{equation}
with $\mu(\phi)$ denoting the product density of hidden variables and $p_1(\phi|\psi)$ the product HV-density conditioned on a single jump. Then,
\begin{equation}\begin{aligned}
&\bar{f}(\psi)=\alpha f(\psi)\\
&+\omega\sum_{s\ge 0}\left((1-\sigma)(1-\omega)\right)^s\left(1-(1-\sigma)^s\right)\int_0^1 f(\phi) p_1(\phi|\psi)d\phi\\
&+\omega\sum_{s\ge 0}(1-\omega)^s\left(1-(1-\sigma)^s\right)^2\int_0^1f(\phi)\mu(\phi)d\phi.\\
\end{aligned}\end{equation}
Note that $\int_0^1f(\phi)\mu(\phi)d\phi=\langle f\rangle=\langle k\rangle/n$. Then, evaluating sums,
\begin{equation}\label{eq:eff_con_pro_product}
\bar{f}(\psi)=\alpha_2 f(\psi)+(\alpha_1-\alpha_2)f_1(\psi)+(1-2\alpha_1+\alpha_2)\langle f\rangle,
\end{equation}
with $\alpha_b=\omega/(1-(1-\omega)(1-\sigma)^b)$, and the quantity $f_1(\psi)$ being defined as
\begin{equation}
f_1(\psi)=\int f(\phi)p_1(\phi|\psi)d\phi,
\end{equation}
where $p_1(\phi|\psi)$ is the distribution of the product of a uniform random variable and of one factor of a product, given that the value of that product is $\psi$. In the following, we walk through the remaining required calculations to obtain $f_1(\psi)$.

\subsection{Finding $p(x|xy=\psi)$}
\begin{figure}
\includegraphics[scale=0.68]{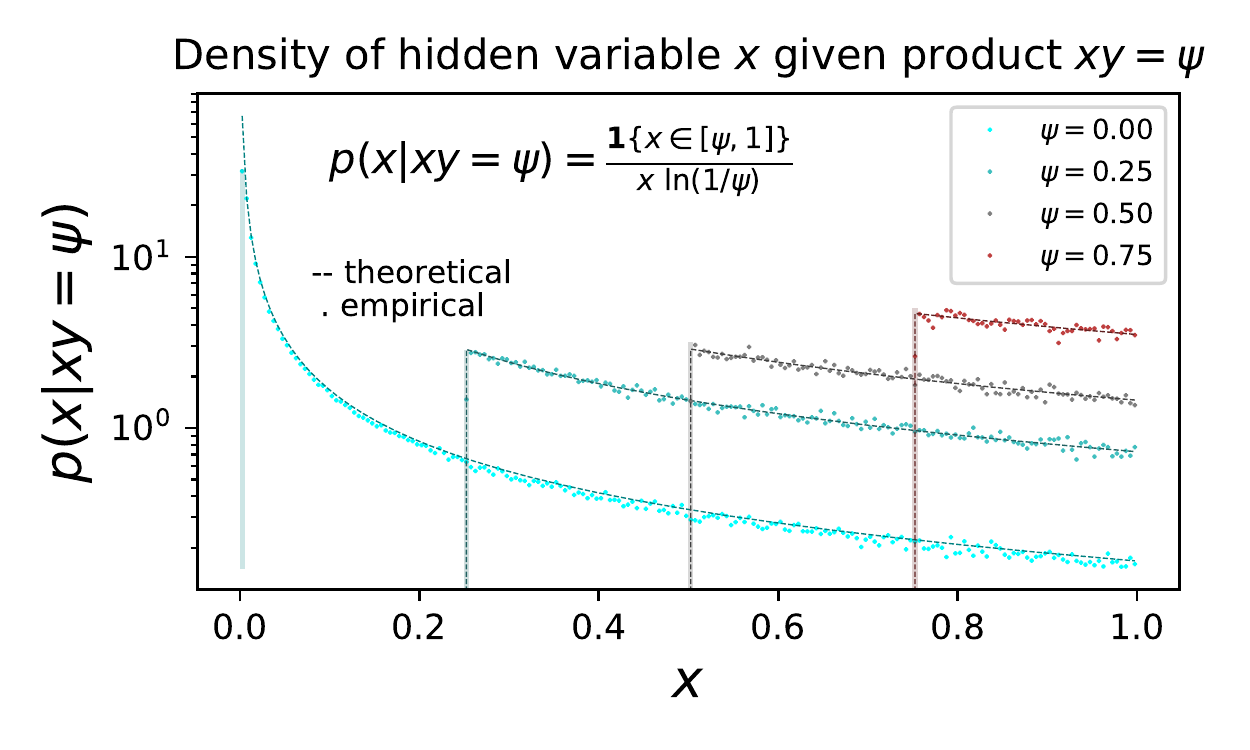}
\caption{The probability density of the value of one member $x$ of a product $xy$ conditioned on $xy=\psi$. In the absence of the conditionality, both $x$ and $y$ are distributed uniformly on $[0,1]$.}
\label{fig:product_conditional_density}
\end{figure}

Suppose that $x$ and $y$ are sampled uniformly on $[0,1]$. Now condition on the fact that their product, $xy$, takes on the particular value $xy=\psi$. Then, what is the probability density of $x$ alone? Note first that it must reside in $[\psi,1]$, since $\psi$ is the product of two numbers each in the range $[0,1]$, {\it i.e.}, each reducing the value of the product. Within the acceptable range, the density is obtained as follows:
\begin{equation}\begin{aligned}
p(x|xy=\psi)&\propto \int_0^1 \mathbf{1}_{\psi}(xy)dy\\
&\propto\frac{1}{x}\int_0^1 \mathbf{1}_{\psi/x}(y)dy\\
&=1/x,
\end{aligned}\end{equation}
where the ratio $\psi/x$ is guaranteed to be in the range $[0,1]$ since $x\ge \psi$. Combining the above with the range of acceptable values of $x$ given $xy=\psi$, we have proportionality
\begin{equation}
p(x|xy=\psi)=c\frac{\mathbf{1}\{x\in[\psi,1]\}}{x},
\end{equation} 
and the normalizing coefficient $c$ is determined by integration:
\begin{equation}\begin{aligned}
1&=\int_0^1p(x|xy=\psi)dx=c\int_\psi^1\frac{dx}{x}=c\ln(1/\psi)\\
&\Rightarrow c=\frac{1}{\ln(1/\psi)}.
\end{aligned}\end{equation}
Therefore,
\begin{equation}
p(x|xy=\psi)=\frac{\mathbf{1}\{x\in[\psi,1]\}}{x  \ln (1/\psi)},
\end{equation} 
as is confirmed numerically in Figure \ref{fig:product_conditional_density}.

\subsection{Finding $p_1(\phi|\psi)$}
Now suppose that one variable, say $y$, undergoes a random jump ({\it i.e.}, is resampled) and thus becomes a new uniform variable on $[0,1]$. The equality $xy=\psi$ no longer holds, but since it did hold prior to the jump, the variable $x$ remains distributed according to $p(x|xy=\psi)$. Therefore the {\it new} product's value, which we denote by $\phi=xy'$ (where $y'$ is the post-jump version of $y$), has a density $p_1(\phi|\psi)$ of the following form:
\begin{equation}\begin{aligned}
p_1(\phi|\psi)&=\int_0^1 \mathbf{1}\{y'\in[0,1]\}p\left(\left.\frac{\phi}{y'}\right\vert xy=\psi\right)\frac{1}{y'}dy'\\
&=\int_0^1\frac{\mathbf{1}\{\phi/y'\in[\psi,1]\}}{(\phi/y')\ln(1/\psi)}\frac{dy'}{y'}\\
&=\frac{1}{\phi \ln(1/\psi)}\int_0^1\mathbf{1}\{\phi/y'\in[\psi,1]\}dy'.
\end{aligned}\end{equation}

Continuing with a change of variables,
\begin{equation}\begin{aligned}
p_1(\phi|\psi)&=\frac{1}{\ln(1/\psi)}\int_{0}^{1/\phi}\mathbf{1}\{y'/\phi\in[1,1/\psi]\}d(y'/\phi)\\
&=\frac{\min(1/\phi,1/\psi)-1}{\ln(1/\psi)}.
\end{aligned}\end{equation}

\begin{figure}
\includegraphics[scale=0.68]{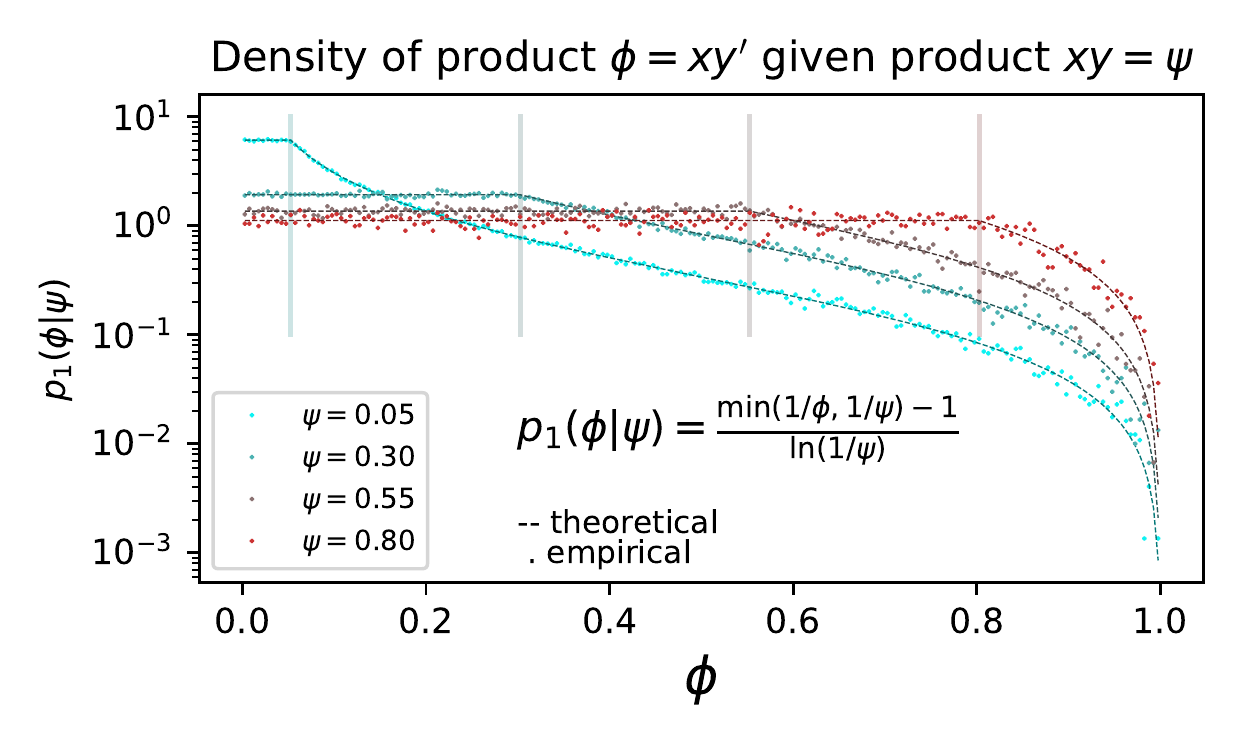}
\caption{The probability density of the value $\phi$ of the product $\phi=xy'$, where $y'$ is uniformly sampled after having previously had random value $y$, and where $xy$ was conditioned to have value $xy=\psi$. Without any conditioning, all three $x,y,y'$ have marginal density uniform on $[0,1]$.}
\label{fig:product_density_after_jump}
\end{figure}

The above is validated numerically in Figure \ref{fig:product_density_after_jump}.

\subsection{Calculating $\bar{f}_1(\psi)$}
We now average the affinity over $p_1(\phi|\psi)$, to get the contribution to the effective connection probability coming from {\it one} hidden variable jumping. This goes as
\begin{equation}\label{eq:barf1}\begin{aligned}
\bar{f}_1(\psi)&=\int_0^1f(\phi)p_1(\phi|\psi)d\phi \\
&=\int_0^1 f(\phi) \frac{\min(1/\phi,1/\psi)-1}{\ln(1/\psi)}d\phi\\
&=\frac{1}{\ln(1/\psi)}\left(\frac{1}{\psi}\int_0^\psi f(\phi)d\phi+\int_\psi^1 \frac{f(\phi)}{\phi}d\phi - 1\right).
\end{aligned}\end{equation}

Using Equation \ref{eq:barf1}, we can compute the effective connection probability $\bar{f}(\psi)$ for a given affinity-function $f(\psi)$ via Equation \ref{eq:eff_con_pro_product}.

\section{Walk-Dynamics}
\label{app:simulating_walk_dynamics}
Here we describe walk-dynamics in detail. Throughout this work, walk-dynamics in $1D$ is simulated by first mapping random variables to the unit interval (by the inverse-CDF method \cite{devroye2006nonuniform}), doing a random walk on $[0,1]$, then mapping back. For any one-dimensional probability density $\nu(x)$ where $x\in\mathbb{R}_+$, we define random variable $u(x)=F_{\nu}(x)$, where $F_\nu(x)=\int_0^x \nu(y)dy$. The probability density of $u(x)$ is the uniform on $[0,1]$. A random walk on $[0,1]$ is constructed via addition of uniform noise in the range $[-2\sigma,2\sigma]$, parameterized by $\sigma\in[0,1]$. That is, after rescaling we have hidden-variable dynamics
\begin{equation}
\mathcal{P}_h(u'|u)=\frac{\mathbf{1}\left\{|u'-u|\le 2\sigma\right\}}{4\sigma}.
\end{equation}
Note that the choice of $[-2\sigma,2\sigma]$ results in a mean jump-length parameterized by $\sigma$:
\begin{equation}\begin{aligned}
\langle |u-u'|\rangle &= \int  |u-u'|\mathcal{P}_h(u'|u)du'\\
&=\frac{1}{4\sigma}\int_{-2\sigma}^{2\sigma}|x|dx=\sigma\\
\end{aligned}\end{equation}
where boundary conditions have been neglected in the above case; when implementing boundary conditions, one needs only to adjust the probability density $\mathcal{P}_h(u'|u)$ according to the circumstance. See Appendix \ref{app:reflecting} for the case of reflecting boundaries.

Drawing $h^{(1)}$ from $\nu$, we initialize $u^{(1)}=F_{\nu}(h^{(1)})$ and iteratively time-advance as per the above to obtain $\{u^{(t)}\}_{t=1}^{T}$. We then simply transform back via $h^{(t)}=F_{\nu}^{-1}(u^{(t)})$, to obtain one-dimensional dynamics whose stationary distribution is $\nu$.

In dimensions greater than $1$, walk-dynamics can be simulated by first taking the multidimensional inverse-CDF transform, mapping the space $\mathcal{X}$ to a unit cube. Walk-dynamics can then be performed with whatever custom boundary conditions are required on that unit cube (boundary conditions that correspond to those of $\mathcal{X}$), and the results can then be mapped back to the original space $\mathcal{X}$. For example in the $\mathbb{H}^2$ model (Appendix \ref{app:H2_walk_dynamics}), increments of change in the angular and radial coordinates were chosen to be independent; this option was taken for simplicity, but non-independent cases would also be interesting to explore. Any transitional probability density preserving the uniform on the unit cube would fall within the same framework.

\section{Walk-dynamics with reflecting boundary conditions}
\label{app:reflecting}
In this appendix we study walk-dynamics on $\mathcal{X}=[0,1]$ with {\it reflecting} boundary conditions under uniform noise. In particular, we show that the stationary density is uniform on $[0,1]$. In turn, that implies that arbitrary 1D dynamics with density $\nu(h)$ can be made into a random walk of this type, by mapping initial $h$-values to $[0,1]$ via the inverse-CDF transform, performing the reflecting random walk on $[0,1]$, then transforming the random walk trajectories back to the original space (see appendix \ref{app:simulating_walk_dynamics}).

Let $\mathcal{X}=[0,1]$ be the HV-space and denote by $x\hookleftarrow U[0,1]$ the value of a hidden variable. Then let $\hat{x}\in[-r,1+r]$ be an intermediate variable defined as $\hat{x}=x+u$, where $u\hookleftarrow U[-r,r]$ is the uniform additive noise which we use to simulate walk-dynamics. Lastly, let $x'=Z(\hat{x})$ be the reflected variable, where the function $Z$ encodes the reflecting boundary conditions. Note that values of $x'$ in the ranges $[0,r]$ and $[1-r,1]$ are obtained from one of two different of values of $\hat{x}$: the case when reflected, and the case when not reflected. To transform the density of $\hat{x}$ into that of $x'$, we write $x'$ as a function of $\hat{z}$, as $x'=Z(\hat{x})$ and use the generalized change-of-variables formula for probability densities \cite{billingsley2008probability}. We denote the densities of $x,\hat{x},x'$ as $P(x),\hat{P}(\hat{x})$, and $P'(x')$, respectively. Then we have the following:
\begin{equation}
P(x')=\sum_{\hat{x}:Z(\hat{x})=x'}\left\vert\frac{d Z(\hat{x})}{d\hat{x}}\right\vert^{-1} P(\hat{x}),
\end{equation}
by the rules of how probabilities densities transform \cite{billingsley2008probability}. The density of $\hat{x}$ given $x$ is
\begin{equation}
\hat{P}(\hat{x}|x)=\frac{\mathbf{1}\{\hat{x}\in[x-r,x+r]\}}{2r}=\frac{\mathbf{1}\{x\in[\hat{x}-r,\hat{x}+r]\}}{2r}.
\end{equation}
Since $x$ is uniform on $[0,1]$ the density of $\hat{x}$ is then
\begin{equation}\begin{aligned}
\hat{P}(\hat{x})&=\int_0^1\hat{P}(\hat{x}|x)dx\\
&=\frac{1}{2r}\int_0^1\mathbf{1}\{x\in[\hat{x}-r,\hat{x}+r]\}dx\\
&=\frac{1}{2r}\left\vert [0,1]\cup[\hat{x}-r,\hat{x}+r]\right\vert,
\end{aligned}\end{equation}
or
\begin{equation}\begin{aligned}
\hat{P}(\hat{x})=\frac{1}{2r}\left\{\begin{array}{cc} \hat{x}+r, & \hat{x}<r, \\ 2r & \hat{x}\in[r,1-r], \\ 1+r-\hat{x}, & \hat{x}>1-r.\end{array}\right.
\end{aligned}\end{equation}
where the $\hat{x}$-dependent coefficients of the first and third terms arise from reflections of the form $\hat{x}-(-r)$ and $1-(\hat{x}-1)$. We seek a function $Z:[-r,1+r]\rightarrow [0,1]$ that encodes the reflection properties of the walk-dynamics. The necessary $Z$ is given by
\begin{equation}\begin{aligned}
Z(\hat{x})&=\left\{\begin{array}{cc} -\hat{x}, & \hat{x}<0, \\ \hat{x}, & \hat{x}\in[0,1], \\ 2-\hat{x}, & \hat{x}>1.\end{array}\right.
\end{aligned}\end{equation}
The values of $\hat{x}$ mapping to a given value of $x'$, namely those making up the inverse of $Z$, are given by
\begin{equation}\begin{aligned}
\{\hat{x}:Z(\hat{x})=x'\}=
\left\{\begin{array}{cc}
 \{-x',x'\}, & x'<r, \\ 
\{x'\}, & x'\in[r,1-r], \\ 
 \{x',2-x'\}, & x'>1-r. \\ 
  \end{array}\right.
\end{aligned}\end{equation}
Let us compute the derivative of $Z(\hat{x})$, neglecting the measure-zero points of $0$ and $1$:
\begin{equation}
\frac{dZ(\hat{x})}{d\hat{x}}=\left\{\begin{array}{cc} -1, & \hat{x}<0, \\ 1, & \hat{x}\in(0,1), \\ -1, & \hat{x}>1.\end{array}\right.
\end{equation}
We now transform to find the density after one step of dynamics, as
\begin{equation}\begin{aligned}
P'(x')&=\sum_{\hat{x}:Z(\hat{x})=x'}\left\vert\frac{d Z(\hat{x})}{d\hat{x}}\right\vert^{-1} \hat{P}(\hat{x})\\
&=
\left\{\begin{array}{cc}
\hat{P}(-x')+\hat{P}(x'), & x'<r, \\ 
\hat{P}(x'), & x'\in[r,1-r], \\ 
 \hat{P}(x')+\hat{P}(2-x'), & x'>1-r. \\ 
  \end{array}\right.\\
&=
\frac{1}{2r}\left\{\begin{array}{cc}
 2r, & x'<r, \\ 
2r, & x'\in[r,1-r], \\ 
2r, & x'>1-r. \\ 
  \end{array}\right.\\
  &=1.
\end{aligned}\end{equation}
Therefore, $P'(x')=1$ for all $x'\in[0,1]$, and thus the uniform distribution is the stationary distribution of reflecting walk-dynamics.

\section{Hyperbolic walk-dynamics}
\label{app:H2_walk_dynamics}
To sample $\tilde{h}=(\tilde{r},\tilde{\theta})$, and also to sample $h_j^{(1)}$ (a coordinate from the initial timestep, {\it i.e.} the static $\mathbb{H}^2$ model), we first draw two independent random variables $U_r$ and $U_\theta$, each from the uniform distribution on $[0,1]$. These are then set equal to the cumulative density functions of $\nu_{rad}$ and $\nu_{ang}$, evaluated at $\tilde{r}$ and $\tilde{\theta}$, respectively:
\begin{equation}
\begin{aligned}
U_\theta &= \int_0^{\tilde{\theta}}\nu_{ang}(\theta)d\theta = \frac{\tilde{\theta}}{2\pi},\\
U_r &= \int_0^{\tilde{r}}\nu_{rad}(r)dr = \frac{\cosh\left(\frac{\gamma-1}{2}\tilde{r}\right)-1}{\cosh\left(\frac{\gamma-1}{2}R\right)-1}.
\end{aligned}
\end{equation}
From the above, we can solve to obtain $\tilde{h}$ in terms of $(U_r,U_\theta)$:
\begin{equation}
\begin{aligned}
\tilde{\theta} &= 2\pi U_\theta,\\
\tilde{r} &= \frac{2}{\gamma-1}\cosh^{-1}\left(1+\left(\cosh\left(\frac{\gamma-1}{2}R\right)-1\right)U_r\right).
\end{aligned}
\end{equation}

In the temporal setting, those initial variables are set to $U^{(1)}_\theta$ and $U^{(1)}_r$, after which we perform walk dynamics on the transformed variables to obtain $(U^{(t)}_{\theta},U^{(t)}_r)$ for $t\in\{2,...,T\}$. Walk-dynamics occurs independently for the two variables, with periodic boundary conditions for angular coordinates and reflecting boundary conditions for radial coordinates. 

Note that we use reflecting boundary conditions for the radial coordinate, rather than, for example, periodic boundary conditions, or reflecting boundary conditions with an associated angular reversal at any timestep that a node reflects from the origin of the radial coordinate (as would also seem like a natural choice for the disk). The reason to not incorporate such angular flipping is due to the interpretation of the angular coordinates as similarity-encoding variables \cite{papadopoulos2012popularity}. From that perspective, it is more realistic to have nodes reflect off of the disk's origin and retain their similarity-coordinates, rather than to pass through the origin and reverse their similarity-coordinates.

\section{Stationarity with Link-Response}
\label{app:stationarity_with_link_response}
In this appendix, we show that the static-model graph probability distribution is preserved via the effect of link-response as described in Section \ref{sec:link_response}. Specifically, we show that 
\begin{equation}\label{eq:stationarity_with_link_response}
\int_{\mathcal{H}}\left(\sum_{G\in\mathcal{G}}\mathbb{P}(G|H)P^{G\rightarrow G'}_{H,H'}\right)\rho(H)dH =\mathbb{P}(G'|H'),
\end{equation}
where $P^{G^{(t)}\rightarrow G^{(t+1)}}_{H^{(t)},H^{(t+1)}}=\mathcal{P}_G(G^{(t+1)}|G^{(t)},H^{(t+1)},H^{(t)})$. We for now set $\omega=0$ and later argue that link-resampling does not influence the results in question. First, we note that the transition probability given $(H,H')$ is separable: $P^{G\rightarrow G'}_{H,H'}=\prod_{1\le i<j\le n}P_{ij}^{A_{ij}\rightarrow A_{ij}'}$, with transition probability $P_{ij}^{\alpha\rightarrow \beta}=\mathbb{P}(A_{ij}'=\beta |A_{ij}=\alpha,h_i',h_j',h_i,h_j)$. Denoting $f_{ij}=f(h_i,h_j)$ and $f'_{ij}=f(h'_i,h'_j)$, we evaluate the different transition probabilities:
\begin{equation}\label{eq:transition_probabilities_link_response}
\begin{aligned}
P_{ij}^{1\rightarrow 1}&=\mathbf{1}\left\{f_{ij}'\ge f_{ij}\right\}+(1-q^-_{ij})\mathbf{1}\{f_{ij}'<f_{ij}\},\\
P_{ij}^{0\rightarrow 0}&=\mathbf{1}\left\{f_{ij}'< f_{ij}\right\}+(1-q^+_{ij})\mathbf{1}\{f_{ij}'\ge f_{ij}\},\\
\end{aligned}
\end{equation}
with $q^-_{ij}=1-f'_{ij}/f_{ij}$, $q^+_{ij}=1-(1-f'_{ij})/(1-f_{ij})$ as defined in Section \ref{sec:link_response}. The remaining probabilities are obtained by normalization:
\begin{equation}\label{eq:transition_probabilities_2_link_response}
\begin{aligned}
P_{ij}^{1\rightarrow 0}&=1-P_{ij}^{1\rightarrow 1}=q^-_{ij}\mathbf{1}\left\{f_{ij}'< f_{ij}\right\},\\
P_{ij}^{0\rightarrow 1}&=1-P_{ij}^{0\rightarrow 0}=q^+_{ij}\mathbf{1}\left\{f_{ij}'\ge f_{ij}\right\}.
\end{aligned}
\end{equation}
\ \\
Noting that $P^{G\rightarrow G'}_{H,H'}$ and $\mathbb{P}(G|H)$ are both separable into a product over $ij:1\le i<j\le n$, we write
\begin{equation}
\mathbb{P}(G|H)P^{G\rightarrow G'}_{H,H'}=\prod_{1\le i<j\le n}f_{ij}^{A_{ij}}(1-f_{ij})^{1-A_{ij}}P_{ij}^{A_{ij}\rightarrow A_{ij}'}.
\end{equation}
The sum over all graphs $G$ of this product becomes a product over all pairs $ij$ of a sum over $A_{ij}\in\{0,1\}$:
\begin{equation}
\sum_{G\in\mathcal{G}}\prod_{1\le i<j\le n}y(A_{ij})=\prod_{1\le i<j\le n}\sum_{A_{ij}\in\{0,1\}}y(A_{ij}).
\end{equation}
Using the above, and the static-model graph probability distribution (Equation \ref{eq:cond_prob}), the parenthesized term in Equation \ref{eq:stationarity_with_link_response} is equal to
\begin{equation}
\begin{aligned}
&\prod_{ij:A_{ij}'=0}\left(f_{ij}P_{ij}^{1\rightarrow 0}+(1-f_{ij})P_{ij}^{0\rightarrow 0}\right) \\
\times& \prod_{ij:A_{ij}'=1} \left(f_{ij}P_{ij}^{1\rightarrow 1}+(1-f_{ij})P_{ij}^{0\rightarrow 1}\right).
\end{aligned}
\end{equation}
Applying equations (\ref{eq:transition_probabilities_link_response}, \ref{eq:transition_probabilities_2_link_response}) and using the expressions for $q^{\pm}_{ij}$, as well as the facts that $\mathbf{1}\{f'_{ij}\ge f_{ij}\}+\mathbf{1}\{f'_{ij}<f_{ij}\}=1$ and $\int_{\mathcal{H}}\rho(H)dH=1$, Equation \ref{eq:stationarity_with_link_response} becomes
\begin{equation}
\prod_{ij:A_{ij}'=1}f'_{ij}\prod_{ij:A_{ij}'=0}(1-f'_{ij})=\mathbb{P}(G'|H').
\end{equation}
The left-hand side of the above is exactly the static model's graph probability distribution given a hidden-variable configuration (see Equation \ref{eq:cond_prob} of the main text). Thus the static hidden-variables model is the stationary distribution of time-advancements with the link-response mechanism.

To show that these results hold even upon inclusion of the link-resampling mechanism (allowing $\omega>0$), consider the following reasoning. Regardless of what the link-response step yielded, each node-pair undergoing link-resampling at rate $\omega$ will result in either (a) linking according to the connection probability of the newly updated hidden-variable configuration (with probability $\omega$) or (b) linking as before ($\omega=0$), without altering the connection probability. Given the fact that stationarity holds without link-resampling, in the latter case we also have a connection probability equal to that of the updated hidden-variable configuration.

Thus, upon inclusion of the link-response mechanism whereby both $H^{(t+1)}$ and $H^{(t)}$ impact the transition from $G^{(t)}$ to $G^{(t+1)}$, we have temporal extensions of arbitrary static hidden-variables models that {\it exactly} satisfy the Equilibrium property, while retaining the Persistence Property. Such a link-response mechanism may better reflect reality in cases where connectivity among nodes changes directly in response to changes in their internal characteristics. 

\section{Discrete hidden variables}
\label{app:discrete_hidden_variables}
We consider THVMs formulated with discrete hidden variables, and describe their relation to continous-HV models. 

We take for example the case of SBMs, described in Section \ref{ssec:temporal_stochastic_block} entirely in terms of discrete HVs, namely group-indices which are naturally thought of as discrete. We then have a set of discrete HVs $\{q_j\}_{j\in [n]}$, each distributed into a discrete set $[m]=\{1,...,m\}$ according to a probability distribution $\varrho:[m]\rightarrow[0,1]$, and connecting via a discrete affinity-function $f_{q,q'}$.

In a dual continuous-HV system which maps to the above-described discrete system, suppose each node $j$'s hidden variable $h_j$ has uniform density on $[0,1]$, and pairwise affinities are encoded in a piecewise constant graphon function according to occupancy of points in nonoverlapping subregions $\{L_w\}_{w\in[m]}\subseteq[0,1]^m$ such that $|L_w|=\varrho_w$ and such that
\begin{equation}
f(h,h')=\sum_{(w,z)\in[m]^2}f_{w,z}\mathbf{1}\{h\in L_w,h'\in L_z\}.
\end{equation}
Discrete node-labels can also be written directly in terms of continuous HV-values, as 
\begin{equation}
q_i=\sum_{q\in[m]}q\mathbf{1}\{h_i\in L_q\}.
\end{equation}
The probability distribution $\varrho$ thus arises from integration of the uniform density on $[0,1]$, namely $\nu(h)=1$, over the regions $\{L_q\}_{q\in[m]}$ corresponding to specific group-labels $q\in[m]$:
\begin{equation}
\varrho_q=\int_{\mathcal{X}}\nu(h)\mathbf{1}\{h\in L_q\}dh=\int_{L_q}dh=|L_q|.
\end{equation}

In the temporal setting, we can again relate discrete HVs to continuous ones. To reproduce the HV-resampling dynamics for temporal SBMs, we can simply have continuous HVs undergo jump-dynamics in $[0,1]$. Jumping to a random point in $[0,1]$ amounts to jumping into a random subset $L_q$ with probability $\varrho_q=|L_q|$.

\bibstyle{apsrev4-1}

\end{document}